\documentclass[twocolumn,eqsecnum,showpacs,showkeys,amsmath,amssymb,nofootinbib,superscriptaddress,floatfix]{revtex4-1}
\usepackage{graphicx}
\usepackage{subfigure}
\usepackage{bm}

\makeatletter
\newcommand\erfc{\mathop{\operator@font erfc}\nolimits}
\def\slashchar#1{\setbox0=\hbox{$#1$}
   \dimen0=\wd0 \setbox1=\hbox{/} \dimen1=\wd1
   \ifdim\dimen0>\dimen1 \rlap{\hbox to \dimen0{\hfil/\hfil}} #1
   \else  \rlap{\hbox to \dimen1{\hfil$#1$\hfil}} / \fi}

\makeatother

\begin{document}
\title{Flow and interferometry in $(3+1)$-dimensional viscous hydrodynamics}
\author{Piotr Bo\.zek}
\email{Piotr.Bozek@ifj.edu.pl}
\affiliation{The H. Niewodnicza\'nski Institute of Nuclear Physics,
PL-31342 Krak\'ow, Poland} \affiliation{
Institute of Physics, Rzesz\'ow University, 
PL-35959 Rzesz\'ow, Poland}
\date{\today}

\begin{abstract}
The  expansion of the fireball created in Au-Au collisions at 
$\sqrt{s_{NN}}=200$GeV is described in $(3+1)$-dimensional viscous 
hydrodynamics with shear and bulk viscosities. We present results for the
     transverse momentum spectra, the directed and elliptic flow and  the
interferometry radii.
\end{abstract}

\pacs{25.75.-q, 25.75.Dw, 25.75.Ld}

\keywords{relativistic 
heavy-ion collisions, viscous  hydrodynamic model, collective flow}

\maketitle

\section{Introduction}

Heavy-ion collisions at ultrarelativistic energies performed at 
 the BNL Relativistic Heavy Ion Collider (RHIC)  
and the CERN Large Hadron Collider (LHC)
\cite{Arsene:2004fa,*Back:2004je,*Adams:2005dq,*Adcox:2004mh,*Aamodt:2010pa,*Aamodt:2011mr,*Aad:2010bu,*Chatrchyan:2011sx} have shown 
that dense matter is formed in the interaction region. The fireball expands 
and a sizable collective  flow develops. 
Effects of the flow are observed in particle spectra, elliptic flow, and 
interferometry radii. Nuclear modification of high $p_\perp$ particle spectra 
 is understood as  the energy loss of partons in the dense medium. 

The dynamics of the dense and hot matter can be quantitatively 
described in terms of relativistic hydrodynamics 
\cite{Kolb:2003dz,*Huovinen:2006jp,*Florkowski:2010zz}. A further refinement 
of the hydrodynamic 
approach involves a finite shear viscosity of the fluid
\cite{Luzum:2008cw,Chaudhuri:2006jd,Song:2007fn,Dusling:2007gi,Bozek:2009dw,Schenke:2010rr}. Finite shear viscosity reduces
the elliptic flow in the system. The comparison of the experimental data
to model predictions for the elliptic flow could be used to estimate the value
 of the shear viscosity coefficient. 
Most of the relativistic viscous hydrodynamic calculations 
for heavy-ion collisions are done in 
$(2+1)$-dimensions [$(2+1)$-D]. Such a simplification requires the 
assumption of boost-invariance of the matter created in the collision.
Experimental data on particle spectra at RHIC show that no boost-invariant 
region  is formed, even for central rapidities
 \cite{Bearden:2004yx}. 
Only recently, first results from a
 full $(3+1)$-D viscous hydrodynamic code have become available 
\cite{Schenke:2010rr}.

We present the results of a relativistic viscous $(3+1)$-D code with shear 
and bulk viscosities applied to Au-Au collisions at $\sqrt{s_{NN}}=200$GeV.
Hydrodynamic calculations are preformed starting from Glauber model
 initial conditions, with the 
freeze-out at $135$MeV and subsequent resonance decays. The use of a realistic 
bulk viscosity in the hadronic phase allows to lower the acceptable 
freeze-out temperature, improving the agreement of the spectra of pions, 
kaons and protons and of the  Hanbury Brown-Twiss  (HBT) correlation 
radii with the data. A low value of the shear viscosity  to entropy ratio
 $\eta/s=0.08$ is consistent with the observed elliptic flow.
The expansion with finite viscosity yields HBT radii closer to the data than 
from ideal fluid hydrodynamics. We present first results on the directed flow
in $(3+1)$-D viscous hydrodynamics.

\section{Viscous hydrodynamics}

\label{sec:visc}

The relativistic second order viscous hydrodynamics \cite{IS} 
is based on the extension of the 
energy-momentum tensor of the perfect fluid
\begin{equation}
T^{\mu\nu}_0=(\epsilon+p)u^\mu u^\nu-p g^{\mu\nu}
\label{eq:ideal}
\end{equation}
by the stress corrections from shear $\pi$  and bulk $\Pi$ viscosities
\begin{equation}
T^{\mu\nu}=T^{\mu\nu}_0+\pi^{\mu\nu}+\Pi \Delta^{\mu\nu}  \ .
\label{eq:tv}
\end{equation}
The fluid energy density, 
pressure and four velocity are denoted by 
$\epsilon$, $p$ and $u^\mu$ respectively. 
 The viscous corrections are solutions of the  dynamical equations 
\begin{equation}
\Delta^{\mu \alpha} \Delta^{\nu \beta} u^\gamma \partial_\gamma 
\pi_{\alpha\beta}=\frac{2\eta \sigma^{\mu\nu}-\pi^{\mu\nu}}{\tau_{\pi}}
-\frac{4}{3}\pi^{\mu\nu}\partial_\alpha u^\alpha
\label{eq:pidyn}
\end{equation}
and
\begin{equation}
 u^\gamma \partial_\gamma \Pi=
\frac{-\zeta \partial_\gamma u^\gamma-\Pi}{\tau_{\Pi}}
-\frac{4}{3}\Pi\partial_\alpha u^\alpha  \ .
\label{eq:budyn}
 \end{equation}
$\Delta^{\mu\nu}=g^{\mu\nu}-u^\mu u^\nu$, 
\begin{equation}
\sigma_{\mu\nu}=\frac{1}{2}\left( \nabla_\mu  u_\nu
+\nabla_\mu   u_\nu    -\frac{2}{3}\Delta_{\mu    \nu  }\partial_\alpha
   u^\alpha\right)\ ,
\end{equation}
and
$\nabla^\mu=\Delta^{\mu\nu} \partial_\nu$.
We take for the  relaxation time $\tau_\pi=\frac{3\eta}{T s}$, and assume
$\tau_\Pi=\tau_\pi$.

\begin{figure}
\includegraphics[width=.45\textwidth]{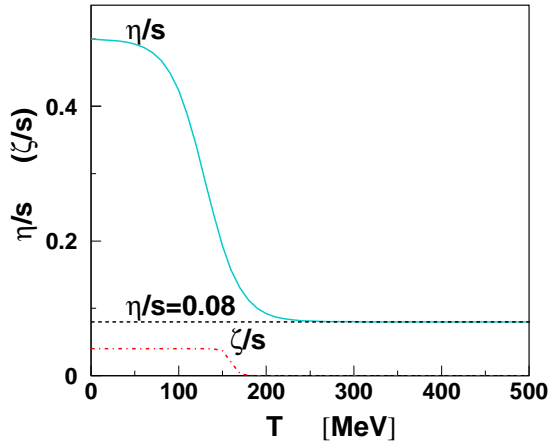}
\caption{(Color online) Temperature dependence of the ratio of shear 
and bulk viscosities to  the entropy. The solid line represents the default 
value, corresponding to $\eta/s=0.08$ in the QGP phase and increasing
 in the hadronic phase to $0.26$ at the  freeze-out, the dashed line 
corresponds to a constant value of $\eta/s=0.08$, and the
 dashed-dotted line represents the bulk viscosity $\zeta/s=0.04$ 
in the hadronic phase.}
\label{fig:etas}
\end{figure}

 Hydrodynamic simulations show that 
the average value of $\eta/s$ must be small in order to describe 
the experimental data on elliptic and triangular flows
\cite{Luzum:2008cw,Shen:2010uy,Alver:2010dn,Schenke:2010rr}. The extracted 
value is close to the conjectured lower limit  $\eta/s=0.08$ 
\cite{Kovtun:2004de} if the Glauber model is used for the initial
profile of the fireball.  

The shear viscosity coefficient
 to entropy ratio is not constant in 
 our default calculations. One expects  significant 
 dissipation and effective viscosity in the hadronic cascade in the 
 last stage of the collision \cite{Hirano:2005xf,Werner:2009fa,*Song:2010aq}.
In the hydrodynamic model, without a hadronic cascade afterburner, it would 
mean that in the hadronic phase $\eta/s$ increases 
\cite{Bozek:2009dw,Niemi:2011ix}.
The viscosity to entropy ratio is taken in the form
\begin{equation}
\frac{\eta}{s}(T)=\frac{\eta_{HG}}{s} f_{HG}(T)+(1-f_{HG}(T))\frac{\eta_{QGP}}{s}
\label{eq:etas}
\end{equation}
with $\eta_{HG}/s=0.5$, $\eta_{QGP}/s=0.08$, and  $f_{HG}(T)=1/\left(\exp\left((T-T_{HG})/\Delta T\right)+1\right)$, where $T_{HG}=130$MeV, $\Delta T=30$MeV.
Bulk viscosity is expected to be negligible in the high temperature plasma
 phase and it must be finite  in the interacting gas of massive hadrons
\cite{NoronhaHostler:2008ju,*Dobado:2009ek,*Demir:2008tr}.
We put a nonzero bulk viscosity coefficient in the hadronic phase
\begin{equation}
\frac{\zeta}{s}(T)=\frac{\zeta_{HG}}{s} f_{\zeta}(T)
\label{eq:zetas}
\end{equation}
with $\zeta_{HG}/s=0.04$ and  $f_{\zeta}(T)=1/\left(\exp\left((T-T_{\zeta})/\Delta T_\zeta\right)+1\right)$, where $T_{\zeta}=160$MeV, $\Delta T_\zeta=4$MeV.
A similar value of the bulk viscosity and of its temperature dependence has been used in the description of the RHIC and LHC 
data with $(2+1)$-D viscous hydrodynamics 
\cite{Bozek:2009dw,Bozek:2010er,Bozek:2011wa}.
The temperature dependence of the viscosity coefficients is shown in Fig. 
\ref{fig:etas}. At the freeze-out temperature $T_f=135$MeV we have 
$\eta/s=0.26$ and $\zeta/s=0.04$.

\begin{figure}
\includegraphics[width=.45\textwidth]{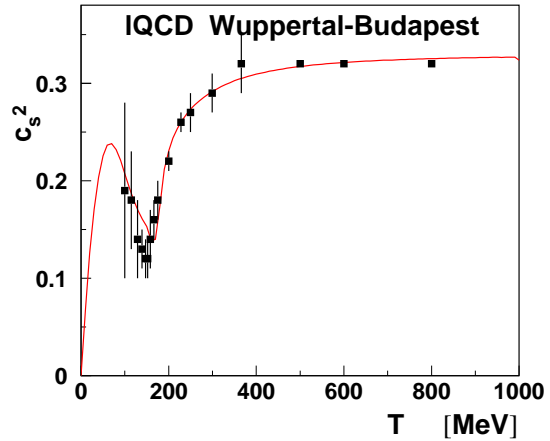}
\caption{(Color online) Temperature dependence of the velocity of 
sound squared used in the hydrodynamic calculations, compared to lattice 
QCD results of \cite{Borsanyi:2010cj}.}
\label{fig:cs}
\end{figure}

\begin{figure}
\includegraphics[width=.45\textwidth]{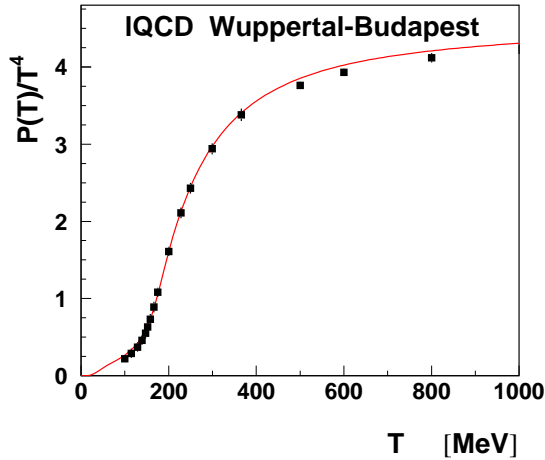}
\caption{(Color online) Temperature dependence of the  pressure $P/T^4$,
 compared to lattice 
QCD results of \cite{Borsanyi:2010cj}. }
\label{fig:eos}
\end{figure}

The equation of state relating the thermodynamical quantities 
is a necessary ingredient for the hydrodynamical evolution.
In the recent years, it became customary for the 
 hydrodynamical calculations to use 
for the equation of state a parametrization of the lattice QCD data combined 
with a non interacting, hadron resonance gas model at lower temperatures
\cite{Luzum:2008cw,Chojnacki:2007jc,Broniowski:2008vp,Pratt:2008qv,Huovinen:2009yb}.
Such an equation of state with a cross-over transition from the plasma
 to the hadronic phase yields a much better description of the measured 
interferometry radii. The resolution of the HBT puzzle is a strong 
argument in favor of the present quantitative understanding of the 
dense matter equation of state at zero baryon density 
\cite{Broniowski:2008vp,Pratt:2008qv}. In the present paper we follow the
prescription of Ref. \cite{Chojnacki:2007jc}, connecting the velocity
 of sound in the  hadron 
gas below $145$MeV to lattice QCD values above $175$MeV. 
The interpolation between the two limiting forms is  
such  that the entropy  from lattice QCD 
is reproduced at high temperatures
\cite{Chojnacki:2007jc}. We use the recent lattice QCD  results 
of the Wuppertal-Budapest group 
\cite{Borsanyi:2010cj}.
The velocity of sound and pressure as function of  temperature are shown 
in Figs. \ref{fig:cs} and \ref{fig:eos}.

The initial  density profile $\rho(\eta_\parallel,x,y)$ 
 for the  hydrodynamic
 evolution
in the space-time rapidity $\eta_\parallel$ 
and the transverse plane $(x,y)$ at an impact parameter $b$ 
is taken  in the form
\begin{eqnarray}
\rho(b,\eta_\parallel,x,y) &=&  \frac{(y_b+\eta_\parallel)N_+
+(y_b-\eta_\parallel)N_-}{y_b(N_++N_-)} 
 \nonumber \\ &&  \left[ \frac{1-\alpha}{2}\rho_{part}
 +\alpha \rho_{bin}\right] f(\eta_\parallel) 
\label{eq:eini}
\end{eqnarray}
and the entropy density is
\begin{equation}
s(\eta_\parallel,x,y)= s_0\frac{\rho(b,\eta_\parallel,x,y)}{\rho(0,0,0,0)} \ ,
\end{equation}
where the density in the transverse plane is proportional to a combination of 
participant nucleon $\rho_{part}=N_++N_-$ and binary collision 
$\rho_{bin}$ densities, with $\alpha=0.125$.
The  densities of the right and left
 going participant 
nucleons $N_\pm(x,y)$ are calculated from the Glauber model.
The first factor on the right hand side of (\ref{eq:eini}) 
gives a tilt of the source away from the beam axis ($y_b$ is
 the beam rapidity). It is motivated by the 
observed asymmetry of the emission of the participant nucleons. A participant
nucleon emits particles preferentially in the same rapidity hemisphere 
\cite{Bialas:2004su}. The hydrodynamic evolution of a tilted source generates the directed flow of particles, as observed in Au-Au collisions at 
$\sqrt{s_{NN}}=200$GeV \cite{Bozek:2010bi}.
In the viscous hydrodynamic calculation we use the participant eccentricity 
for the initial fireball \cite{Alver:2008zz}. 
The optical Glauber model gives the standard 
eccentricity.  We rescale the density in the transverse plane 
($x\rightarrow x/\beta$, $y\rightarrow y\beta$) to  
reproduce the participant eccentricity obtained from a 
 Glauber Monte-Carlo model \cite{Broniowski:2007nz} , with $\beta =1.02$-$1.03$.
The parameters and the longitudinal profile 
\begin{equation}
f(\eta_\parallel)=\exp\left(-\frac{(\eta_\parallel-\eta_0)^2}
{2\sigma_\eta^2}\theta(|\eta_\parallel|-\eta_0)
\right)
\label{eq:etaprofile}
\end{equation}
are adjusted to reproduce the charged particle distribution in pseudorapidity.
We have  $\eta_0=1.5$, $\sigma_\eta=1.4$ for the viscous evolution and
 $\eta_0=1.7$, $\sigma_\eta=1.4$ for the perfect fluid evolution.
The parameters of the Woods-Saxon density distribution in Au nuclei
\begin{equation}
\rho(x,y,z)=\frac{\rho_0}{1+\exp\left((\sqrt{x^2+y^2+z^2}-R_A)/a\right)} \ ,
\end{equation}
are  $\rho_0=0.17 \mbox{fm}^{-3}$, $R_A=6.38$fm and
$a=0.535$fm; 
and the nucleon-nucleon cross section is $42$mb.
For the hydrodynamic evolution starting at $\tau_0=0.6$fm/c,
the maximal entropy density $s_0$ corresponds to a temperature of $380$MeV,
i.e. an energy density of $33$GeV/fm$^3$.

The hydrodynamic equations 
\begin{equation}
\partial_\mu T^{\mu\nu}=0
\end{equation}
together with the equations for the stress corrections (\ref{eq:pidyn})
 and  (\ref{eq:budyn})
are solved numerically in the $x$, $y$, $\eta_\parallel$ coordinates
starting from $\tau_0$. The initial flow is the Bjorken scaling flow 
$u^\mu=(t/\tau,0,0,z/\tau)$,
 the initial shear stress tensor takes the Navier-Stokes form, and 
$\Pi(\tau_0)=0$.
The  evolution in $\tau=\sqrt{t^2-z^2}$ 
is performed in a two-step method with spatial 
gradients calculated on a grid with spacing $\Delta x=\Delta y=0.24$fm.
At small times, viscosity corrections to the pressure are substantial. 
The formalism of second order viscous hydrodynamics is not applicable in that 
case \cite{Huovinen:2008te}.
To regularize $\pi^{\mu\nu}$ we use the formula 
\begin{equation}
\pi^{\mu\nu}_{reg}=\frac{\pi^{\mu\nu}}{\left(1+ 
\frac{4 (\pi^{\alpha \beta}\pi_{\alpha \beta})^2}{9 p^4} \right)^{1/4} },
\end{equation}
assuring that the longitudinal pressure does not get negative, even in the
 early phase of the evolution.

At the freeze-out temperature of $135$MeV and $145$MeV for viscous and
ideal fluid hydrodynamics, particles are emitted from the 
freeze-out hypersurface according to the 
 Cooper-Frye formula.
Viscous corrections to the equilibrium  momentum distribution $f_0$
\begin{equation}
f=f_0+\delta f_{shear}+\delta f_{bulk}
\label{eq:deltaf}
\end{equation}
yield a change in the energy-momentum tensor in hadronic phase
\begin{equation}
T^{\mu\nu}=\sum_n \int \frac{d^3 p}{(2\pi)^3 E } p^\mu p^\nu
\left( f_0 + \delta f  \right) = T^{\mu \nu}_0 + \delta T^{\mu\nu} \end{equation}
the sum is over all the hadron species.
The corrections to the energy-momentum tensor fulfill the Landau matching conditions
\begin{equation}
u_\mu \delta T^{\mu \nu}u_\nu=0
\label{eq:lm1}
\end{equation}
and
\begin{equation}
u_\mu\delta N^\mu_k=0
\label{eq:lm2}
\end{equation}
where 
\begin{equation}
\delta N^\mu =\sum_n \int \frac{d^3 p}{(2\pi)^3 E }   b_k p^\mu  \delta f   
\end{equation}
is the change in the conserved charge $b_k$ (e.g. baryon number, strangeness) in the system.

The form of the stress corrections to the energy-momentum tensor and the matching 
conditions do not determine uniquely the nonequilibrium corrections $\delta f$, 
either in the form of the momentum dependence
 or the contribution of different hadrons in a multicomponent system.
  We use a  quadratic form for  the shear viscosity corrections 
\begin{equation}
\delta f_{shear}= f_0
\left(1\pm f_0 \right) \frac{1}{2 T^2 (\epsilon+p)}p^\mu p^\nu \pi_{\mu\nu}
\label{eq:dfsh}
\end{equation}
and an  asymptotically linear 
form for the bulk viscosity, based on the relaxation time approximation 
\cite{Gavin:1985ph,*Hosoya:1983xm,*Sasaki:2008fg,Bozek:2009dw},
\begin{equation}
\delta f_{bulk}= C_{bulk}f_0
\left(1\pm f_0 \right)\left(c_s^2 u^\mu p_\mu -\frac{(u^\mu p_\mu)^2-m^2}
{3 u^\mu p_\mu}\right) \Pi  \ ,
\end{equation}
with, in the local rest frame,
\begin{equation}
\frac{1}{C_{bulk}}= \frac{1}{3}\sum_n\int \frac{d^3 p}{(2\pi)^3}\frac{m^2}{E}f_0
\left(1\pm f_0 \right)\left(c_s^2 E -\frac{p^2}{3 E}\right) \ .
\label{eq:dfbu}
\end{equation} 

The form of  the shear viscosity corrections is standard and commonly used
 \cite{Teaney:2003kp}, on the other hand, 
different expressions for the bulk viscosity corrections are considered, Grad expansion 
\cite{Monnai:2009ad}, exponential \cite{Pratt:2010jt}, and the relaxation 
time formula 
\cite{Bozek:2009dw,Gavin:1985ph,Dusling:2011fd}. The assumed
 form and the relative contribution 
of different hadron species to the bulk viscosity 
corrections are important as  they  influence
 the transverse momentum spectra of produced
particles
and their relative yields.
In general the constraints imposed by the Landau matching conditions
 lead to chemical nonequilibrium corrections from the bulk viscosity. 
It is true for any form of the assumed momentum dependence  of the bulk viscosity 
corrections, linear, quadratic 
or exponential. Imposing chemical equilibrium would require Landau matching conditions not
for the conserved quantum numbers, but separately for each particle specie.
The actual  choice of the bulk viscosity correction to be used needs a specific assumption 
on the reequilibration rates for different particles.
The proposed forms of the bulk viscosity corrections range from assuming a similar form
for all hadrons \cite{Monnai:2009ad,Bozek:2009dw,Gavin:1985ph}, to a  different  form for 
 mesons, baryons and strange particles \cite{Dusling:2011fd}, to requiring the  particle 
numbers to be unchanged \cite{Pratt:2010jt}.
We use  formula (\ref{eq:dfbu})
 assuming   a common  
relaxation time for all the particles, which leads to deviations from 
chemical equilibrium due to bulk viscosity corrections.
 It is a minimal assumption, 
but more elaborate ansatze are possible with different relaxation times for different 
particles for both bulk  and shear viscosity 
corrections \cite{Dusling:2009df,*Molnar:2011kx,Dusling:2011fd}.
The modification of the momentum distribution (\ref{eq:dfbu}) fulfills the Landau matching
conditions due to the relation
\begin{eqnarray} & & 
\sum_n \int \frac{d^3p}{(2\pi)^3 }  C_{bulk} f_0
\left(1\pm f_0 \right)\left(c_s^2 E^2 -\frac{p^2}
{3 }\right) \Pi  \nonumber \\
& &  =0  \ .
\end{eqnarray}

To illustrate the effects of chemical nonequilibrium, 
let us consider a system undergoing a fast, Hubble-like expansion with a
collective velocity
$u^\mu=x^\mu/\tau_3$, $\tau_3=\sqrt{t^2-x^2-y^2-z^2}$. If at some proper time $\tau_3=t_0$ the 
interactions are turned off,
 the particle distributions at later times are
\begin{equation}
f(p,x)=f_0(\sqrt{m^2+(p \tau_3/t_0)^2}/T_{dec}) \ ,
\label{eq:vla}
\end{equation}
if at $\tau_3=t_0$ the momentum distributions $f_0(E/T_{dec})$ 
are in   equilibrium 
 at the temperature $T_{dec}$.
The distributions  (\ref{eq:vla}) are solutions of the Vlasov equation
\begin{equation}
p^\mu\partial_\mu f(p,x)=0 \ .
\end{equation} 

At times $\tau_3>t_0$ the
 energy density
\begin{equation}
\epsilon(\tau_3)=\sum_n \int \frac{d^3p}{(2\pi)^3}E f_n(p,\tau_3) 
\label{eq:enhubb}
\end{equation}
drops.
In a real system  rescatterings after $t_0$ 
are still present and the particle distribution
is driven toward the equilibrium with the temperature $T_{eq}$ corresponding to the energy density 
$\epsilon_{eq}=\epsilon(\tau_3)$. This equilibration is incomplete, and one can use an ansatz for the 
distribution with bulk corrections at the freeze-out of  the form
\begin{equation}
f_0+\delta f_{bulk}=f_0(\sqrt{m^2+p^2 \lambda}/T_{eff})
\label{eq:bexp}
\end{equation}
with the parameters $T_{eff}$ and $\lambda$ adjusted to reproduce 
the matching conditions
\begin{equation}
\sum_n \int \frac{d^3p}{(2\pi )^3} E f_0(\sqrt{m^2+p^2 \lambda}/T_{eff})=\epsilon_{eq}
\end{equation}
and
\begin{equation}
\sum_n \int \frac{d^3p}{(2\pi )^3E} \frac{p^2}{3} f_0(\sqrt{m^2+p^2 \lambda}/T_{eff})=p_{eq}+\Pi \ .
\end{equation}
The freeze-out temperature $T_{eq}$ corresponds to the energy 
density $\epsilon_{eq}$, but the 
particle ratios correspond to 
the temperature $T_{eff}$ with  $T_{eq}<T_{eff}<T_{dec}$. The equilibration 
processes drive the momentum distribution 
function from the distribution (\ref{eq:vla}) towards the equilibrium one 
with the temperature $T_{eq}$. The ansatz (\ref{eq:bexp})
 describes this effect of partial chemical 
and kinetic reequilibration after $t_0$. Using the simple 
two-parameter formula (\ref{eq:bexp})
the effect of the deviations from 
equilibrium are taken into account. If
 the chemical reequilibration processes are significantly slower 
than the kinetic ones, the particle
 ratios get fixed at some chemical freeze-out temperature $T_{ch}$ 
with $T_{eff}<T_{ch}<T_{dec}$. 
Even in that case  using $T_{eff}$, instead 
of forcing the particle ratios to be fixed at the temperature given by the 
energy density $T_{eq}$, reduces the error.

  The difference between $T_{eff}$ and the true chemical freeze-out
temperature $T_{ch}$ is not big if reequilibration processes 
are defined by the energy scales,
which means that it is
 as difficult to repopulate a pion state with momentum 
$800$MeV as an $\omega$ state 
with momentum  $220$MeV. The momentum distribution with bulk viscosity 
corrections (\ref{eq:bexp}) is in chemical equilibrium at the temperature
 $T_{eff}$. But  because we compare it to the reference 
equilibrium distributions 
corresponding to the temperature $T_{eq}$ given by the energy density, 
the particle ratios are off equilibrium for $T_{eq}$, the reason is that 
equilibration processes are not fast enough to repopulate high 
momentum states and depopulate  high mass states relative to the 
instantaneous, approximate 
equilibrium state defined by the Landau matching condition. 
Extensive calculation in $(2+1)$-D viscous hydrodynamics
with the bulk viscosity corrections of the form (\ref{eq:bexp}) 
or (\ref{eq:dfbu}) 
give almost indistinguishable results for the final
 spectra \cite{Bozek:2010er,Bozek:2011wa,Bozek:2011ph}. 
In both cases, due to the  shift in the temperature 
from $T_{eq}$ to $T_{eff}$ 
the particle ratios appear as off chemical equilibrium for the freeze-out
temperature $T_{eq}$.

Particle emission and resonance decay is performed using the 
Monte-Carlo generator THERMINATOR2
\cite{Chojnacki:2011hb}. The hydrodynamic expansion is done 
using the equation of state at zero 
baryon density. At RHIC in the central rapidity region, the baryon chemical potential is nonzero
 yielding  for the ratio of antiprotons to protons $\simeq 0.8$. We reintroduce the 
nonzero chemical potentials at the freeze-out with the ratio $\mu/T$ taken 
 from thermal model fits 
\cite{Andronic:2005yp}. This procedure violates the baryon number flow at the freeze-out hypersurface, 
and approximately is equivalent to multiplying the final proton
 spectra by $\exp(\mu/T)$ and the antiproton spectra by $\exp(-\mu/T)$. 
The justification for 
this procedure is that the equation of state is expected be moderately 
changing with $\mu$ 
at small baryon densities \cite{Bluhm:2007nu} and that 
the energy is conserved at the freeze-out to the order $\mu^2/T^2$. 
The net effect is mainly the rescaling of the relative numbers of protons
 and antiprotons, which is crucial for comparing with experimental spectra 
at central rapidities.

\section{Results}

\begin{figure}
\includegraphics[width=.5\textwidth]{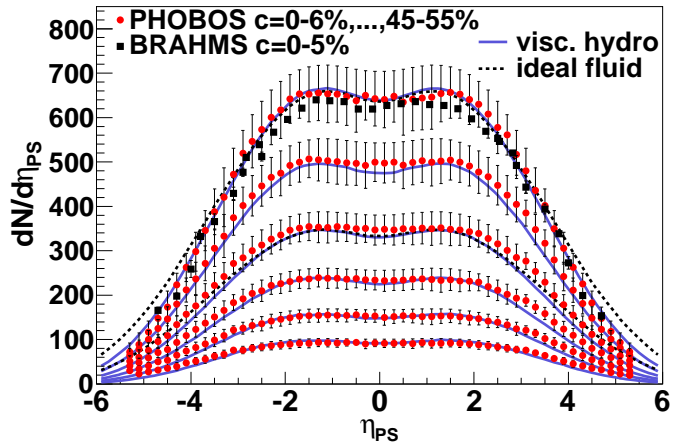}
\caption{(Color online) Pseudorapidity distribution 
of charged hadrons for centrality classes $0-6\%$, $6-15\%$, $15-25\%$, 
$25-35\%$, $35-45\%$ and $45-55\%$ calculated in the viscous and perfect fluid hydrodynamics (solid and dashed 
lines respectively) compared to PHOBOS Collaboration data (dots) 
\cite{Back:2002wb}. The squares represent the BRAHMS Collaboration 
data for centrality $0-5\%$ 
  \cite{Bearden:2001qq}.}
\label{fig:dndeta}
\end{figure}

The distribution of charged particles in pseudorapidity 
is shown in Fig. \ref{fig:dndeta} for different centralities.
The width of the initial distribution of  matter for the hydrodynamic 
evolution in Eq. (\ref{eq:eini}) is adjusted to reproduce the final 
charged hadron distribution. It is interesting to compare the parameters for
 the viscous and perfect fluid evolutions. The initial 
width for the viscous hydrodynamics is smaller.  A similar behavior of 
the matter distribution in the
longitudinal direction in the
$(3+1)$-D evolution has been observed in Ref. \cite{Schenke:2011bn}. 
It is contrary to the 
expectations from  simple $1+1$D viscous hydrodynamic calculations
\cite{Bozek:2007qt}.
The reduced longitudinal pressure in the initial stage
of the evolution
\begin{equation}
p+\pi^{zz}
\label{eq:pl}
\end{equation}
should lead to a reduced expansion of the matter in space-time rapidity. 
Such a reduced expansion is observed
in $1+1$D calculations, using narrow initial 
distributions in space-time rapidity \cite{Bozek:2007qt,Monnai:2011ju}.
In $(3+1)$-D evolution, the initial distribution has a broad plateau
 in space-time rapidity, where no expansion occurs at early times.
In the tails, outside of the plateau, the expansion is faster in the 
ideal fluid case (see the tails of the distributions in Fig. \ref{fig:dndeta}).

\begin{figure}
\includegraphics[width=.45\textwidth]{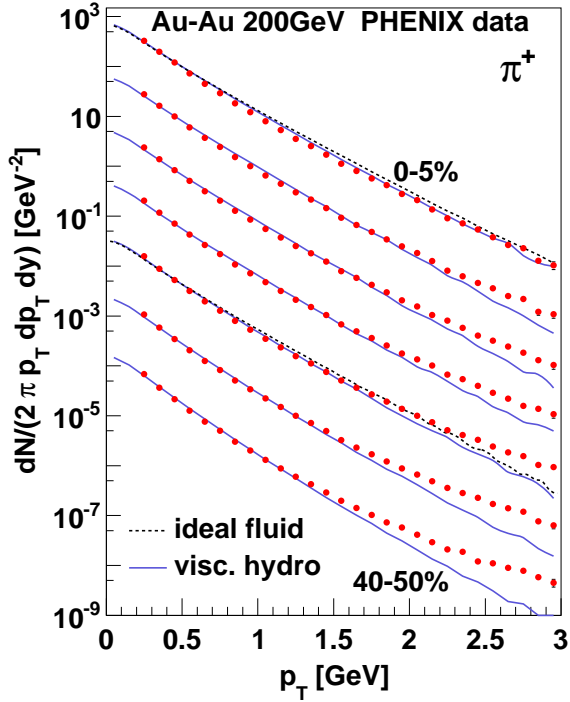}
\caption{(Color online) $\pi^+$ 
transverse momentum spectra  for 
 centralities $0$-$5$\%, $5$-$10$\%, $10$-$15$\%, $15$-$20$\%,
 $20$-$30$\%, $30$-$40$\% and $40$-$50$\% (successively scaled down 
by powers of $0.1$) from viscous hydrodynamic calculations (solid lines). 
The dashed lines for the 
 centralities $0-5\%$ and $20-30\%$ represent the results of 
the perfect fluid hydrodynamics. 
Data are from the PHENIX collaboration \cite{Adler:2003cb}. }
\label{fig:ptpion}
\end{figure}

\begin{figure}
\includegraphics[width=.45\textwidth]{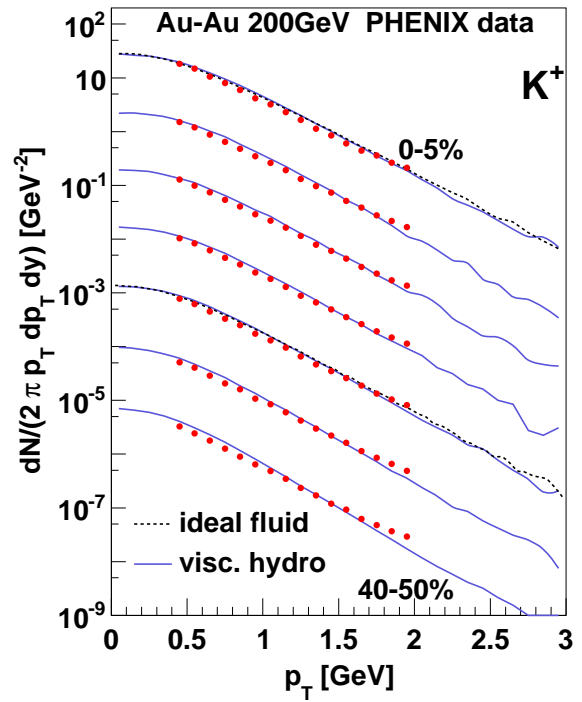}
\caption{(Color online) The same as Fig \ref{fig:ptpion} but for $K^+$. }
\label{fig:ptkaon}
\end{figure}

\begin{figure}
\includegraphics[width=.45\textwidth]{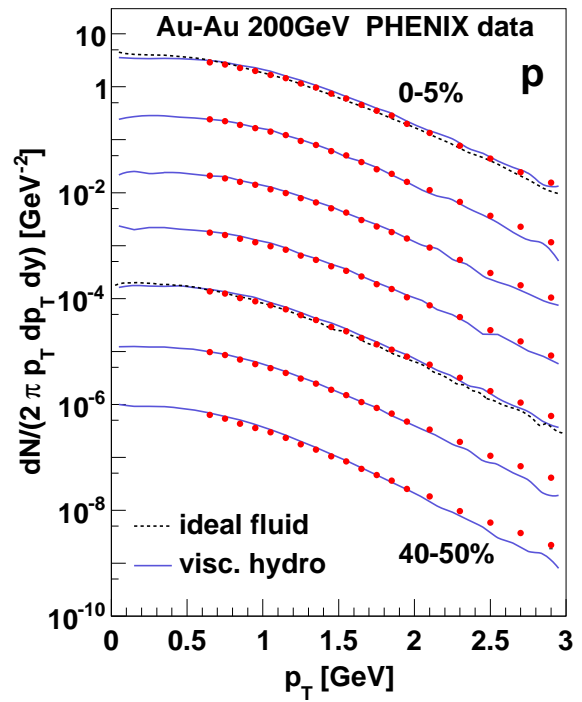}
\caption{(Color online) The same as Fig \ref{fig:ptpion} but for protons. }
\label{fig:ptproton}
\end{figure}

The centrality dependence of $dN/d\eta_{PS}$ is reproduced using the
initial entropy density scaled with a combination of participant nucleons 
and binary collisions. The parameter  $\alpha=0.125$ for the 
admixture of binary collisions is smaller than seen in the 
final density at $\eta_{PS}=0$ ($\alpha=0.145$ \cite{Back:2004dy}).
The difference comes from the interplay of the longitudinal 
and transverse expansions, and from the entropy production in the
viscous hydrodynamics.

Pion spectra in transverse momentum are well reproduced at different
 centralities (Fig. \ref{fig:ptpion}), for $p_\perp<1.2$GeV.
The role of the bulk viscosity is essential in reproducing the spectra, 
as it
reduces the effective thermal motion of the emitted particles. The collective 
component is larger, corresponding to a lower freeze-out temperature.
Very similar
 results are obtained using ideal hydrodynamics, but at a higher
 freeze-out temperature $145$MeV. It means that the $p_\perp$ distributions 
in the perfect fluid case are obtained using a smaller collective flow, 
but larger thermal motion.

\begin{figure}
\includegraphics[width=.45\textwidth]{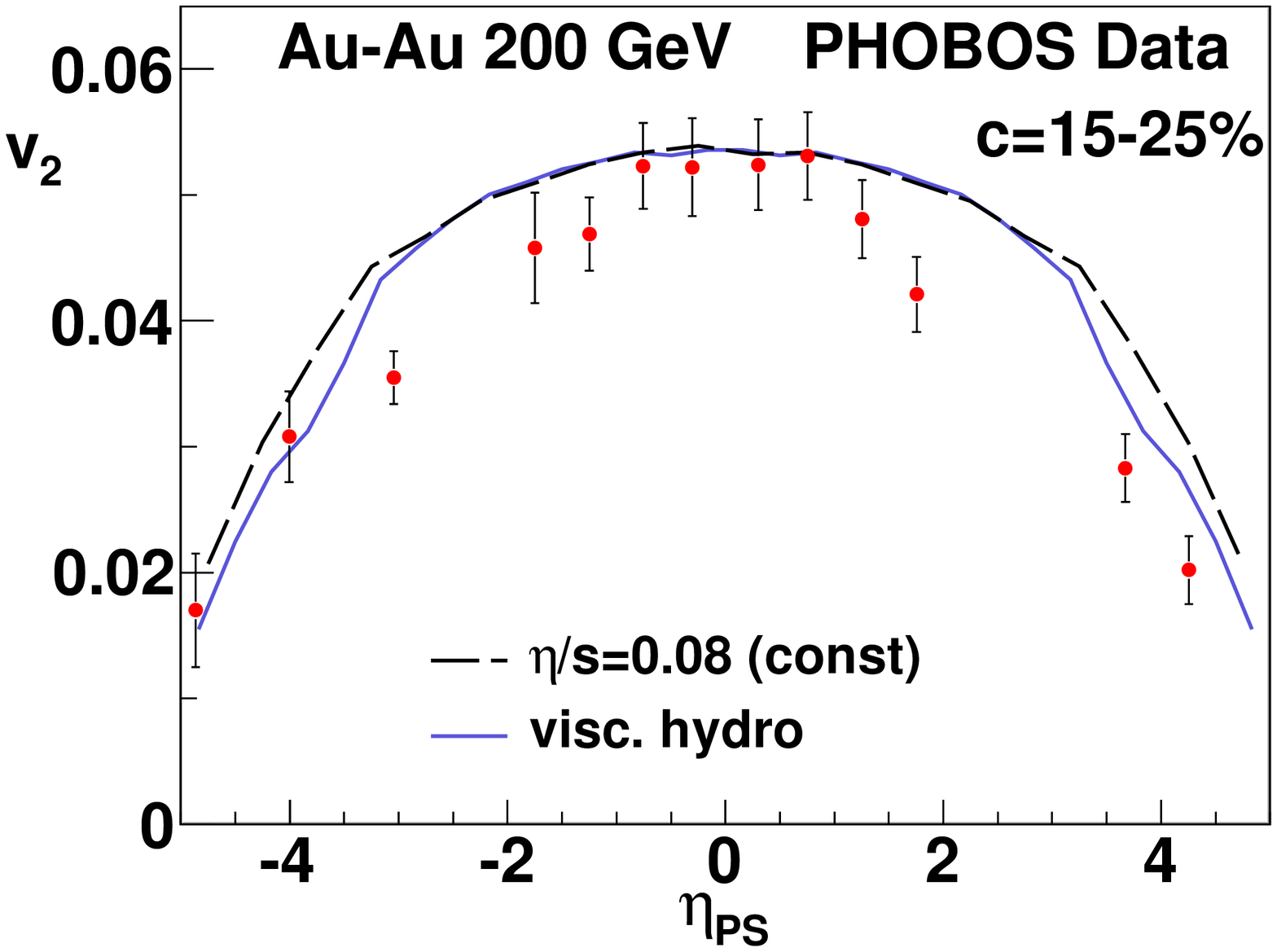}
\caption{(Color online) Pseudorapidity dependence of the 
elliptic flow coefficient for charged particles 
for centralities  $15-25\%$, for the viscous hydrodynamic expansion with 
increasing (solid line) and constant (dashed line) shear viscosity to entropy
in the hadronic phase,  data from the PHOBOS collaboration
 are denoted by dots
\cite{Back:2004mh}. }
\label{fig:v2eta}
\end{figure}

Kaon spectra are well reproduced in central collisions (Fig. \ref{fig:ptkaon}).
In semi-peripheral collisions  the number of kaons is
overpredicted. It may be a sign of the incomplete equilibration of 
strangeness in 
peripheral collisions
\cite{Cleymans:2004pp,*Bozek:2005eu,*Becattini:2008ya}. The same as for pions, 
the perfect and viscous fluid hydrodynamics give similar results.
Hydrodynamic calculations describe the proton spectra up to $p_\perp<2$GeV 
(Fig. \ref{fig:ptproton}).
Small differences can be observed between perfect fluid and viscous 
calculations. Viscosity leads to harder spectra for protons, as 
heavy particles are more sensitive to the collective flow. 
Another difference is that the number of protons is larger in the viscous
 calculations, although the freeze-out temperature is lower, and one would 
expect a smaller thermal rate of production. This is an effect of the bulk 
viscosity, that drives the system out of chemical equilibrium. In an expanding
system with bulk viscosity the ratio of the number  of heavy to light particles 
is larger than predicted in chemical equilibrium at $T_f$ 
(Sect. \ref{sec:visc}).
The spectra for pions, kaons,  and protons are very similar as obtained 
in $(2+1)$-D viscous hydrodynamics with bulk and shear viscosity 
\cite{Bozek:2009dw}.

\begin{figure}
\includegraphics[width=.48\textwidth]{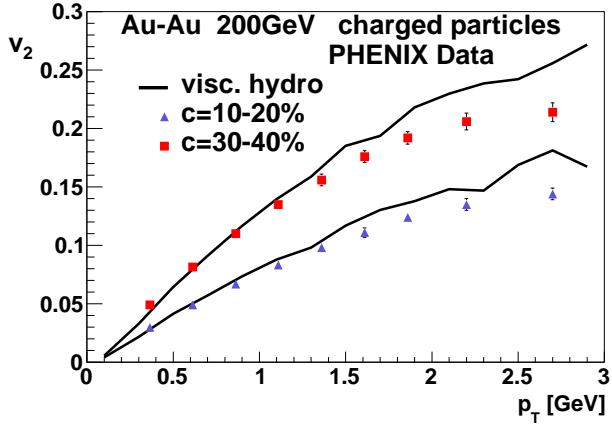}

\caption{(Color online) Elliptic flow of charged particles as function of 
transverse momentum from viscous hydrodynamic calculations 
 for centralities $c=10-20\%$ (lower line) and $c=30-40\%$ 
(upper line) compared to PHENIX collaboration data \cite{Adare:2011tg}.  }
\label{fig:v2ch10}
\end{figure}

\begin{figure}
\includegraphics[width=.48\textwidth]{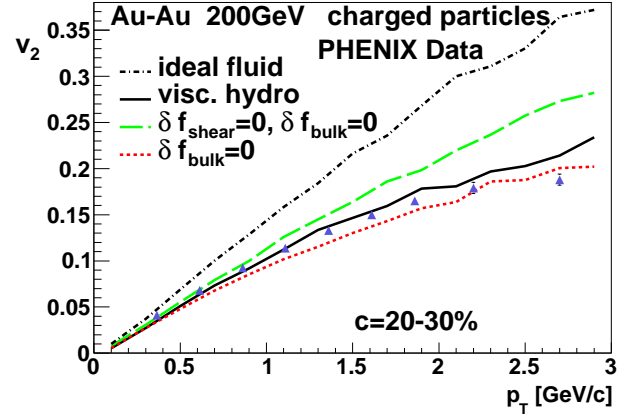}

\caption{(Color online) Elliptic flow of charged particles as function of 
transverse momentum from viscous hydrodynamic calculations 
 for centralities $c=20-30\%$ (solid line), from 
perfect fluid hydrodynamics (dashed-dotted line), from the viscous hydrodynamics without $\delta f_{shear}$ and $\delta f_{bulk}$  
corrections at the freeze-out (dashed line), and from the viscous hydrodynamics without $\delta f_{bulk}$  
corrections at the freeze-out (dotted line). Data are from the 
 PHENIX collaboration \cite{Adare:2011tg}.  }
\label{fig:v2chpt}
\end{figure}

\begin{figure}
\includegraphics[width=.5\textwidth]{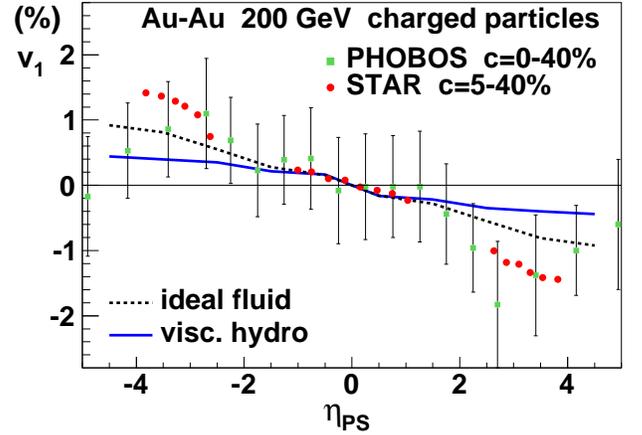}
\caption{(Color online) Directed flow in Au-Au 
  collisions,  perfect fluid (dotted line) and viscous 
fluid hydrodynamics (solid line) are
 compared to experimental data from the PHOBOS and STAR collaborations
\cite{Back:2005pc,Abelev:2008jga}. }
\label{fig:v1}
\end{figure}

A characteristic, for which viscous evolution in $(3+1)$-D may be important is 
the pseudorapidity dependence of the elliptic flow. Ideal fluid hydrodynamics
gives a flat dependence, unlike measured in experiments.
Dissipative effects in the hadronic cascade bring the results of the
simulations closer to the  data \cite{Hirano:2005xf}.
 In terms of the viscous hydrodynamics, one expects 
stronger shear viscosity corrections at forward rapidities, where 
the matter freezes-out earlier \cite{Bozek:2009mz}.
$(3+1)$-D viscous hydrodynamic calculations by Schenke et al. give a flat dependence of $v_2$ on the pseudorapidity 
\cite{Schenke:2010rr,Schenke:2011bn}, both using the average
 and  fluctuating initial conditions. The same can be observed in our 
calculation (Fig. \ref{fig:v2eta}). We study the effect of the increase of the
shear viscosity in the hadronic phase (solid line) 
as compared to a calculation
 using a constant $\eta/s$ (dashed line). We observe 
a minor improvement of 
the agreement with the data when the viscosity increases, as would have been
expected 
if the effect determining  the shape of $v_2(\eta_{PS})$
 were the dissipation in  the hadron  cascade; it is not enough to 
remove the discrepancy with the PHOBOS measurements. In the simulation with 
constant $\eta/s$ we start the evolution with the standard eccentricity, 
given by the optical Glauber model. When using fluctuating initial conditions,
 the initial eccentricity is the participant eccentricity, but the expansion 
of lumpy initial conditions reduces the final flow \cite{Schenke:2010rr}. 
These effects depend on the centrality, the coarse-graining of the 
initial fluctuations and on the viscosity \cite{Schenke:2011bn,Qiu:2011fi}.

The elliptic flow of charged particles as function of $p_\perp$
is shown in Fig. \ref{fig:v2ch10} and \ref{fig:v2chpt} for three
 different centrality classes. Viscous hydrodynamics gives
 a satisfactory description for $p_\perp<1.5$GeV. The reduction of the
elliptic flow from viscosity happens in the hydrodynamic phase (dashed-dotted 
line versus dashed line in Fig. \ref{fig:v2chpt}). An additional reduction of 
the azimuthal asymmetry happens due to the shear viscosity corrections at
 freeze-out (dotted line in Fig. \ref{fig:v2chpt}). The inclusion of bulk 
viscosity 
corrections increases the elliptic flow slightly, as noted in 
\cite{Monnai:2009ad}. The reason is that bulk viscosity reduces the thermal 
motion of the emitted particles and the  momenta are more aligned 
with the collective flow of the fluid. The fluid velocity is transverse
 to the shear stress tensor $u^\mu \pi_{\mu\nu}=0$, so the shear 
viscosity correction in Eq. (\ref{eq:dfsh}) is reduced.

\begin{figure}
\includegraphics[width=.35\textwidth]{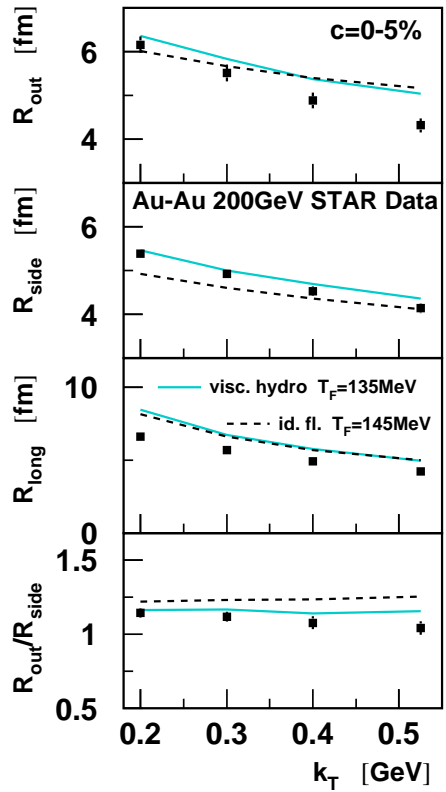}
\caption{(Color online) HBT radii for Au-Au collisions at centralities $0-5\%$.
Ideal fluid  results 
(dashed lines), viscous hydrodynamic  results  (solid lines), 
 and STAR Collaboration data \cite{Adams:2004yc} (squares) are shown. }
\label{fig:hbt}
\end{figure}

The expansion of the tilted source (Eq. \ref{eq:eini}) gives a sizable 
negative directed flow (Fig. \ref{fig:v1}). The perfect fluid dynamics
 gives a larger $v_1$, as  predicted in Ref. \cite{Bozek:2010aj}.
The formation of the directed flow from the tilted source, involves 
the simultaneous action of the transverse and longitudinal pressures
 in the fluid \cite{Bozek:2010bi} and it happens early in the evolution.
Shear viscosity corrections reduce the longitudinal pressure (Eq. \ref{eq:pl}) 
and increase the transverse one, as a result less directed flow is generated.
The directed flow observable is potentially a very sensitive measure of the 
pressure imbalance at the early stage. However,
significant  uncertainties are still present 
related to the initial conditions,
the starting time of the evolution, and the 
nature of the initial pressure imbalance
\cite{Bozek:2007di,*Ryblewski:2010bs,*Martinez:2010sc,*Beuf:2009cx,*Heller:2011ju}.

The HBT radii 
are calculated from the correlations
of identical pions. Correlated pairs are generated from the Monte-Carlo events 
\cite{Chojnacki:2011hb}.
The interferometry radii as a function of the momentum 
pair $k_\perp$ are shown in Fig. \ref{fig:hbt}.
The calculations are in good agreement with STAR data; with viscous 
hydrodynamics being  slightly better. The size of the emitting 
source $R_{side}$ is correctly predicted. The larger collective 
flow in the calculation
involving both the shear and bulk  viscosities  yields a smaller value of
the ratio  $R_{out}/R_{side}$. We find the role of bulk viscosity important 
to obtain a common description of both the HBT radii and the transverse 
momentum spectra. The results for the
 HBT radii in the $(3+1)$-D calculations are very similar as in $(2+1)$-D 
  for both the viscous  \cite{Bozek:2009dw,Bozek:2010er}
and perfect fluid \cite{Bozek:2009ty} dynamics.

\section{Conclusions}

We present a set of $(3+1)$-D viscous hydrodynamic calculations for Au-Au
collisions at $\sqrt{s_{NN}}=200$GeV. Using an independently developed 
hydrodynamic code coupled to a statistical emission and resonance decay 
event generator \cite{Chojnacki:2011hb} we evaluate several soft observables for
final particles. 
The measured spectra of pions, kaons, and protons are well reproduced. The 
elliptic flow of charged particles as a function of $p_\perp$ is  calculated
for different centralities. A small value of the shear viscosity $\eta/s=0.08$
for calculations using Glauber model initial conditions yields results 
 compatible with the data.
 The dependence of the flow coefficient $v_2$ 
on pseudorapidity is too flat, even when taking into account 
the increase of the shear 
viscosity in the hadronic phase. The directed flow of charged particles
is calculated for the
 viscous fluid evolution, finding a reduction 
 compared to the perfect fluid case. The interferometry radii 
as a function of the momentum of the pair of pions are calculated.
Viscosity improves the agreement with the data, especially for $R_{side}$ and 
the ratio $R_{out}/R_{side}$.

The values for some of the observables studied, the transverse momentum 
spectra, the elliptic
 flow as a function of $p_\perp$ and the HBT radii are similar as in  $(2+1)$-D 
viscous hydrodynamic simulations. The elliptic flow as a function of 
pseudorapidity is similar to previous $(3+1)$-D calculations using averaged 
initial conditions \cite{Schenke:2010rr,Schenke:2011bn}. The reduction 
of the directed flow from viscosity comes as expected \cite{Bozek:2010aj}.

\section*{Acknowledgment}
Supported by 
Polish Ministry of Science and Higher Education under
grant N N202 263438.

\bibliography{../hydr}

\begin{thebibliography}{10}%
\makeatletter
\providecommand \@ifxundefined [1]{%
 \ifx #1\undefined \expandafter \@firstoftwo
 \else \expandafter \@secondoftwo
\fi
}%
\providecommand \@ifnum [1]{%
 \ifnum #1\expandafter \@firstoftwo
 \else \expandafter \@secondoftwo
\fi
}%
\providecommand \enquote [1]{``#1''}%
\providecommand \bibnamefont  [1]{#1}%
\providecommand \bibfnamefont [1]{#1}%
\providecommand \citenamefont [1]{#1}%
\providecommand\href[0]{\@sanitize\@href}%
\providecommand\@href[1]{\endgroup\@@startlink{#1}\endgroup\@@href}%
\providecommand\@@href[1]{#1\@@endlink}%
\providecommand \@sanitize [0]{\begingroup\catcode`\&12\catcode`\#12\relax}%
\@ifxundefined \pdfoutput {\@firstoftwo}{%
 \@ifnum{\z@=\pdfoutput}{\@firstoftwo}{\@secondoftwo}%
}{%
 \providecommand\@@startlink[1]{\leavevmode\special{html:<a href="#1">}}%
 \providecommand\@@endlink[0]{\special{html:</a>}}%
}{%
 \providecommand\@@startlink[1]{%
  \leavevmode
  \pdfstartlink
   attr{/Border[0 0 1 ]/H/I/C[0 1 1]}%
   user{/Subtype/Link/A<</Type/Action/S/URI/URI(#1)>>}%
  \relax
 }%
 \providecommand\@@endlink[0]{\pdfendlink}%
}%
\providecommand \url  [0]{\begingroup\@sanitize \@url }%
\providecommand \@url [1]{\endgroup\@href {#1}{\urlprefix}}%
\providecommand \urlprefix [0]{URL }%
\providecommand \Eprint[0]{\href }%
\@ifxundefined \urlstyle {%
  \providecommand \doi [1]{doi:\discretionary{}{}{}#1}%
}{%
  \providecommand \doi [0]{doi:\discretionary{}{}{}\begingroup
  \urlstyle{rm}\Url }%
}%
\providecommand \doibase [0]{http://dx.doi.org/}%
\providecommand \Doi[1]{\href{\doibase#1}}%
\providecommand \bibAnnote [3]{%
  \BibitemShut{#1}%
  \begin{quotation}\noindent
    \textsc{Key:}\ #2\\\textsc{Annotation:}\ #3%
  \end{quotation}%
}%
\providecommand \bibAnnoteFile [2]{%
  \IfFileExists{#2}{\bibAnnote {#1} {#2} {\input{#2}}}{}%
}%
\providecommand \typeout [0]{\immediate \write \m@ne }%
\providecommand \selectlanguage [0]{\@gobble}%
\providecommand \bibinfo [0]{\@secondoftwo}%
\providecommand \bibfield [0]{\@secondoftwo}%
\providecommand \translation [1]{[#1]}%
\providecommand \BibitemOpen[0]{}%
\providecommand \bibitemStop [0]{}%
\providecommand \bibitemNoStop [0]{.\EOS\space}%
\providecommand \EOS [0]{\spacefactor3000\relax}%
\providecommand \BibitemShut [1]{\csname bibitem#1\endcsname}%
\bibitem{Arsene:2004fa}%
  \BibitemOpen
  \bibfield{author}{%
  \bibinfo {author} {\bibfnamefont{I.}~\bibnamefont{Arsene}} \emph{et~al.}
  (\bibinfo {collaboration} {BRAHMS}),\ }%
  \bibfield{journal}{%
  \Doi{10.1016/j.nuclphysa.2005.02.130}{\bibinfo {journal} {Nucl. Phys.}}\ }%
  \textbf{\bibinfo {volume} {A757}},\ \bibinfo {pages} {1} (\bibinfo {year}
  {2005})%
  \bibAnnoteFile{NoStop}{Arsene:2004fa}%
\bibitem{Back:2004je}%
  \BibitemOpen
  \bibfield{author}{%
  \bibinfo {author} {\bibfnamefont{B.~B.}\ \bibnamefont{Back}} \emph{et~al.}
  (\bibinfo {collaboration} {PHOBOS}),\ }%
  \bibfield{journal}{%
  \Doi{10.1016/j.nuclphysa.2005.03.084}{\bibinfo {journal} {Nucl. Phys.}}\ }%
  \textbf{\bibinfo {volume} {A757}},\ \bibinfo {pages} {28} (\bibinfo {year}
  {2005})%
  \bibAnnoteFile{NoStop}{Back:2004je}%
\bibitem{Adams:2005dq}%
  \BibitemOpen
  \bibfield{author}{%
  \bibinfo {author} {\bibfnamefont{J.}~\bibnamefont{Adams}} \emph{et~al.}
  (\bibinfo {collaboration} {STAR}),\ }%
  \bibfield{journal}{%
  \Doi{10.1016/j.nuclphysa.2005.03.085}{\bibinfo {journal} {Nucl. Phys.}}\ }%
  \textbf{\bibinfo {volume} {A757}},\ \bibinfo {pages} {102} (\bibinfo {year}
  {2005})%
  \bibAnnoteFile{NoStop}{Adams:2005dq}%
\bibitem{Adcox:2004mh}%
  \BibitemOpen
  \bibfield{author}{%
  \bibinfo {author} {\bibfnamefont{K.}~\bibnamefont{Adcox}} \emph{et~al.}
  (\bibinfo {collaboration} {PHENIX}),\ }%
  \bibfield{journal}{%
  \Doi{10.1016/j.nuclphysa.2005.03.086}{\bibinfo {journal} {Nucl. Phys.}}\ }%
  \textbf{\bibinfo {volume} {A757}},\ \bibinfo {pages} {184} (\bibinfo {year}
  {2005})%
  \bibAnnoteFile{NoStop}{Adcox:2004mh}%
\bibitem{Aamodt:2010pa}%
  \BibitemOpen
  \bibfield{author}{%
  \bibinfo {author} {\bibfnamefont{K.}~\bibnamefont{Aamodt}} \emph{et~al.}
  (\bibinfo {collaboration} {ALICE}),\ }%
  \bibfield{journal}{%
  \bibinfo {journal} {Phys.Rev.Lett.}\ }%
  \textbf{\bibinfo {volume} {105}},\ \bibinfo {pages} {252302} (\bibinfo {year}
  {2010})%
  \bibAnnoteFile{NoStop}{Aamodt:2010pa}%
\bibitem{Aamodt:2011mr}%
  \BibitemOpen
  \bibfield{author}{%
  \bibinfo {author} {\bibfnamefont{K.}~\bibnamefont{Aamodt}} \emph{et~al.}
  (\bibinfo {collaboration} {ALICE}),\ }%
  \bibfield{journal}{%
  \Doi{10.1016/j.physletb.2010.12.053}{\bibinfo {journal} {Phys. Lett.}}\ }%
  \textbf{\bibinfo {volume} {B696}},\ \bibinfo {pages} {328} (\bibinfo {year}
  {2011})%
  \bibAnnoteFile{NoStop}{Aamodt:2011mr}%
\bibitem{Aad:2010bu}%
  \BibitemOpen
  \bibfield{author}{%
  \bibinfo {author} {\bibfnamefont{G.}~\bibnamefont{Aad}} \emph{et~al.}
  (\bibinfo {collaboration} {Atlas}),\ }%
  \bibfield{journal}{%
  \bibinfo {journal} {Phys. Rev. Lett.}\ }%
  \textbf{\bibinfo {volume} {105}},\ \bibinfo {pages} {252303} (\bibinfo {year}
  {2010})%
  \bibAnnoteFile{NoStop}{Aad:2010bu}%
\bibitem{Chatrchyan:2011sx}%
  \BibitemOpen
  \bibfield{author}{%
  \bibinfo {author} {\bibfnamefont{S.}~\bibnamefont{Chatrchyan}} \emph{et~al.}
  (\bibinfo {collaboration} {CMS}),\ }%
  \bibfield{journal}{%
  \Doi{10.1103/PhysRevC.84.024906}{\bibinfo {journal} {Phys. Rev.}}\ }%
  \textbf{\bibinfo {volume} {C84}},\ \bibinfo {pages} {024906} (\bibinfo {year}
  {2011})%
  \bibAnnoteFile{NoStop}{Chatrchyan:2011sx}%
\bibitem{Kolb:2003dz}%
  \BibitemOpen
  \bibfield{author}{%
  \bibinfo {author} {\bibfnamefont{P.~F.}\ \bibnamefont{Kolb}}\ and\ \bibinfo
  {author} {\bibfnamefont{U.~W.}\ \bibnamefont{Heinz}},\ }%
  in\ \emph{\bibinfo {booktitle} {Quark Gluon Plasma 3}},\ \bibinfo {editor}
  {edited by\ \bibinfo {editor} {\bibfnamefont{R.}~\bibnamefont{Hwa}}\ and\
  \bibinfo {editor} {\bibfnamefont{X.~N.}\ \bibnamefont{Wang}}}\ (\bibinfo
  {publisher} {World Scientific, Singapore},\ \bibinfo {year} {2004})\
  \Eprint{http://arxiv.org/abs/nucl-th/0305084}{arXiv:nucl-th/0305084}%
  \bibAnnoteFile{NoStop}{Kolb:2003dz}%
\bibitem{Huovinen:2006jp}%
  \BibitemOpen
  \bibfield{author}{%
  \bibinfo {author} {\bibfnamefont{P.}~\bibnamefont{Huovinen}}\ and\ \bibinfo
  {author} {\bibfnamefont{P.~V.}\ \bibnamefont{Ruuskanen}},\ }%
  \bibfield{journal}{%
  \Doi{10.1146/annurev.nucl.54.070103.181236}{\bibinfo {journal} {Ann. Rev.
  Nucl. Part. Sci.}}\ }%
  \textbf{\bibinfo {volume} {56}},\ \bibinfo {pages} {163} (\bibinfo {year}
  {2006})%
  \bibAnnoteFile{NoStop}{Huovinen:2006jp}%
\bibitem{Florkowski:2010zz}%
  \BibitemOpen
  \bibfield{author}{%
  \bibinfo {author} {\bibfnamefont{W.}~\bibnamefont{Florkowski}},\ }%
  \emph{\bibinfo {title} {{Phenomenology of Ultra-Relativistic Heavy-Ion
  Collisions}}}\ (\bibinfo {publisher} {World Scientific Publishing Company,
  Singapore},\ \bibinfo {year} {2010})%
  \bibAnnoteFile{NoStop}{Florkowski:2010zz}%
\bibitem{Luzum:2008cw}%
  \BibitemOpen
  \bibfield{author}{%
  \bibinfo {author} {\bibfnamefont{M.}~\bibnamefont{Luzum}}\ and\ \bibinfo
  {author} {\bibfnamefont{P.}~\bibnamefont{Romatschke}},\ }%
  \bibfield{journal}{%
  \Doi{10.1103/PhysRevC.78.034915}{\bibinfo {journal} {Phys. Rev.}}\ }%
  \textbf{\bibinfo {volume} {C78}},\ \bibinfo {pages} {034915} (\bibinfo {year}
  {2008})%
  \bibAnnoteFile{NoStop}{Luzum:2008cw}%
\bibitem{Chaudhuri:2006jd}%
  \BibitemOpen
  \bibfield{author}{%
  \bibinfo {author} {\bibfnamefont{A.~K.}\ \bibnamefont{Chaudhuri}},\ }%
  \bibfield{journal}{%
  \bibinfo {journal} {Phys. Rev.}\ }%
  \textbf{\bibinfo {volume} {C74}},\ \bibinfo {pages} {044904} (\bibinfo {year}
  {2006})%
  \bibAnnoteFile{NoStop}{Chaudhuri:2006jd}%
\bibitem{Song:2007fn}%
  \BibitemOpen
  \bibfield{author}{%
  \bibinfo {author} {\bibfnamefont{H.}~\bibnamefont{Song}}\ and\ \bibinfo
  {author} {\bibfnamefont{U.~W.}\ \bibnamefont{Heinz}},\ }%
  \bibfield{journal}{%
  \Doi{10.1016/j.physletb.2007.11.019}{\bibinfo {journal} {Phys. Lett.}}\ }%
  \textbf{\bibinfo {volume} {B658}},\ \bibinfo {pages} {279} (\bibinfo {year}
  {2008})%
  \bibAnnoteFile{NoStop}{Song:2007fn}%
\bibitem{Dusling:2007gi}%
  \BibitemOpen
  \bibfield{author}{%
  \bibinfo {author} {\bibfnamefont{K.}~\bibnamefont{Dusling}}\ and\ \bibinfo
  {author} {\bibfnamefont{D.}~\bibnamefont{Teaney}},\ }%
  \bibfield{journal}{%
  \Doi{10.1103/PhysRevC.77.034905}{\bibinfo {journal} {Phys. Rev.}}\ }%
  \textbf{\bibinfo {volume} {C77}},\ \bibinfo {pages} {034905} (\bibinfo {year}
  {2008})%
  \bibAnnoteFile{NoStop}{Dusling:2007gi}%
\bibitem{Bozek:2009dw}%
  \BibitemOpen
  \bibfield{author}{%
  \bibinfo {author} {\bibfnamefont{P.}~\bibnamefont{Bo\.zek}},\ }%
  \bibfield{journal}{%
  \bibinfo {journal} {Phys. Rev.}\ }%
  \textbf{\bibinfo {volume} {C81}},\ \bibinfo {pages} {034909} (\bibinfo {year}
  {2010})%
  \bibAnnoteFile{NoStop}{Bozek:2009dw}%
\bibitem{Schenke:2010rr}%
  \BibitemOpen
  \bibfield{author}{%
  \bibinfo {author} {\bibfnamefont{B.}~\bibnamefont{Schenke}}, \bibinfo
  {author} {\bibfnamefont{S.}~\bibnamefont{Jeon}},\ and\ \bibinfo {author}
  {\bibfnamefont{C.}~\bibnamefont{Gale}},\ }%
  \bibfield{journal}{%
  \Doi{10.1103/PhysRevLett.106.042301}{\bibinfo {journal} {Phys. Rev. Lett.}}\
  }%
  \textbf{\bibinfo {volume} {106}},\ \bibinfo {pages} {042301} (\bibinfo {year}
  {2011})%
  \bibAnnoteFile{NoStop}{Schenke:2010rr}%
\bibitem{Bearden:2004yx}%
  \BibitemOpen
  \bibfield{author}{%
  \bibinfo {author} {\bibfnamefont{I.~G.}\ \bibnamefont{Bearden}} \emph{et~al.}
  (\bibinfo {collaboration} {BRAHMS}),\ }%
  \bibfield{journal}{%
  \bibinfo {journal} {Phys. Rev. Lett.}\ }%
  \textbf{\bibinfo {volume} {94}},\ \bibinfo {pages} {162301} (\bibinfo {year}
  {2005})%
  \bibAnnoteFile{NoStop}{Bearden:2004yx}%
\bibitem{IS}%
  \BibitemOpen
  \bibfield{author}{%
  \bibinfo {author} {\bibfnamefont{W.}~\bibnamefont{Israel}}\ and\ \bibinfo
  {author} {\bibfnamefont{J.}~\bibnamefont{Stewart}},\ }%
  \bibfield{journal}{%
  \bibinfo {journal} {Annals Phys.}\ }%
  \textbf{\bibinfo {volume} {118}},\ \bibinfo {pages} {341} (\bibinfo {year}
  {1979})%
  \bibAnnoteFile{NoStop}{IS}%
\bibitem{Shen:2010uy}%
  \BibitemOpen
  \bibfield{author}{%
  \bibinfo {author} {\bibfnamefont{C.}~\bibnamefont{Shen}}, \bibinfo {author}
  {\bibfnamefont{U.}~\bibnamefont{Heinz}}, \bibinfo {author}
  {\bibfnamefont{P.}~\bibnamefont{Huovinen}},\ and\ \bibinfo {author}
  {\bibfnamefont{H.}~\bibnamefont{Song}},\ }%
  \bibfield{journal}{%
  \Doi{10.1103/PhysRevC.82.054904}{\bibinfo {journal} {Phys. Rev.}}\ }%
  \textbf{\bibinfo {volume} {C82}},\ \bibinfo {pages} {054904} (\bibinfo {year}
  {2010})%
  \bibAnnoteFile{NoStop}{Shen:2010uy}%
\bibitem{Alver:2010dn}%
  \BibitemOpen
  \bibfield{author}{%
  \bibinfo {author} {\bibfnamefont{B.~H.}\ \bibnamefont{Alver}}, \bibinfo
  {author} {\bibfnamefont{C.}~\bibnamefont{Gombeaud}}, \bibinfo {author}
  {\bibfnamefont{M.}~\bibnamefont{Luzum}},\ and\ \bibinfo {author}
  {\bibfnamefont{J.-Y.}\ \bibnamefont{Ollitrault}},\ }%
  \bibfield{journal}{%
  \Doi{10.1103/PhysRevC.82.034913}{\bibinfo {journal} {Phys. Rev.}}\ }%
  \textbf{\bibinfo {volume} {C82}},\ \bibinfo {pages} {034913} (\bibinfo {year}
  {2010})%
  \bibAnnoteFile{NoStop}{Alver:2010dn}%
\bibitem{Kovtun:2004de}%
  \BibitemOpen
  \bibfield{author}{%
  \bibinfo {author} {\bibfnamefont{P.~K.}\ \bibnamefont{Kovtun}}, \bibinfo
  {author} {\bibfnamefont{D.~T.}\ \bibnamefont{Son}},\ and\ \bibinfo {author}
  {\bibfnamefont{A.~O.}\ \bibnamefont{Starinets}},\ }%
  \bibfield{journal}{%
  \Doi{10.1103/PhysRevLett.94.111601}{\bibinfo {journal} {Phys. Rev. Lett.}}\
  }%
  \textbf{\bibinfo {volume} {94}},\ \bibinfo {pages} {111601} (\bibinfo {year}
  {2005})%
  \bibAnnoteFile{NoStop}{Kovtun:2004de}%
\bibitem{Hirano:2005xf}%
  \BibitemOpen
  \bibfield{author}{%
  \bibinfo {author} {\bibfnamefont{T.}~\bibnamefont{Hirano}}, \bibinfo {author}
  {\bibfnamefont{U.~W.}\ \bibnamefont{Heinz}}, \bibinfo {author}
  {\bibfnamefont{D.}~\bibnamefont{Kharzeev}}, \bibinfo {author}
  {\bibfnamefont{R.}~\bibnamefont{Lacey}},\ and\ \bibinfo {author}
  {\bibfnamefont{Y.}~\bibnamefont{Nara}},\ }%
  \bibfield{journal}{%
  \Doi{10.1016/j.physletb.2006.03.060}{\bibinfo {journal} {Phys. Lett.}}\ }%
  \textbf{\bibinfo {volume} {B636}},\ \bibinfo {pages} {299} (\bibinfo {year}
  {2006})%
  \bibAnnoteFile{NoStop}{Hirano:2005xf}%
\bibitem{Werner:2009fa}%
  \BibitemOpen
  \bibfield{author}{%
  \bibinfo {author} {\bibfnamefont{K.}~\bibnamefont{Werner}} \emph{et~al.},\ }%
  \bibfield{journal}{%
  \bibinfo {journal} {J. Phys.}\ }%
  \textbf{\bibinfo {volume} {G36}},\ \bibinfo {pages} {064030} (\bibinfo {year}
  {2009})%
  \bibAnnoteFile{NoStop}{Werner:2009fa}%
\bibitem{Song:2010aq}%
  \BibitemOpen
  \bibfield{author}{%
  \bibinfo {author} {\bibfnamefont{H.}~\bibnamefont{Song}}, \bibinfo {author}
  {\bibfnamefont{S.~A.}\ \bibnamefont{Bass}},\ and\ \bibinfo {author}
  {\bibfnamefont{U.}~\bibnamefont{Heinz}},\ }%
  \bibfield{journal}{%
  \Doi{10.1103/PhysRevC.83.024912}{\bibinfo {journal} {Phys.Rev.}}\ }%
  \textbf{\bibinfo {volume} {C83}},\ \bibinfo {pages} {024912} (\bibinfo {year}
  {2011})%
  \bibAnnoteFile{NoStop}{Song:2010aq}%
\bibitem{Niemi:2011ix}%
  \BibitemOpen
  \bibfield{author}{%
  \bibinfo {author} {\bibfnamefont{H.}~\bibnamefont{Niemi}}, \bibinfo {author}
  {\bibfnamefont{G.~S.}\ \bibnamefont{Denicol}}, \bibinfo {author}
  {\bibfnamefont{P.}~\bibnamefont{Huovinen}}, \bibinfo {author}
  {\bibfnamefont{E.}~\bibnamefont{Molnar}},\ and\ \bibinfo {author}
  {\bibfnamefont{D.~H.}\ \bibnamefont{Rischke}},\ }%
  \bibfield{journal}{%
  \Doi{10.1103/PhysRevLett.106.212302}{\bibinfo {journal} {Phys.Rev.Lett.}}\ }%
  \textbf{\bibinfo {volume} {106}},\ \bibinfo {pages} {212302} (\bibinfo {year}
  {2011})%
  \bibAnnoteFile{NoStop}{Niemi:2011ix}%
\bibitem{NoronhaHostler:2008ju}%
  \BibitemOpen
  \bibfield{author}{%
  \bibinfo {author} {\bibfnamefont{J.}~\bibnamefont{Noronha-Hostler}}, \bibinfo
  {author} {\bibfnamefont{J.}~\bibnamefont{Noronha}},\ and\ \bibinfo {author}
  {\bibfnamefont{C.}~\bibnamefont{Greiner}},\ }%
  \bibfield{journal}{%
  \Doi{10.1103/PhysRevLett.103.172302}{\bibinfo {journal} {Phys. Rev. Lett.}}\
  }%
  \textbf{\bibinfo {volume} {103}},\ \bibinfo {pages} {172302} (\bibinfo {year}
  {2009})%
  \bibAnnoteFile{NoStop}{NoronhaHostler:2008ju}%
\bibitem{Dobado:2009ek}%
  \BibitemOpen
  \bibfield{author}{%
  \bibinfo {author} {\bibfnamefont{A.}~\bibnamefont{Dobado}}, \bibinfo {author}
  {\bibfnamefont{F.~J.}\ \bibnamefont{Llanes-Estrada}},\ and\ \bibinfo {author}
  {\bibfnamefont{J.~M.}\ \bibnamefont{Torres-Rincon}},\ }%
  \bibfield{journal}{%
  \Doi{10.1103/PhysRevD.80.114015}{\bibinfo {journal} {Phys. Rev.}}\ }%
  \textbf{\bibinfo {volume} {D80}},\ \bibinfo {pages} {114015} (\bibinfo {year}
  {2009})%
  \bibAnnoteFile{NoStop}{Dobado:2009ek}%
\bibitem{Demir:2008tr}%
  \BibitemOpen
  \bibfield{author}{%
  \bibinfo {author} {\bibfnamefont{N.}~\bibnamefont{Demir}}\ and\ \bibinfo
  {author} {\bibfnamefont{S.~A.}\ \bibnamefont{Bass}},\ }%
  \bibfield{journal}{%
  \Doi{10.1103/PhysRevLett.102.172302}{\bibinfo {journal} {Phys. Rev. Lett.}}\
  }%
  \textbf{\bibinfo {volume} {102}},\ \bibinfo {pages} {172302} (\bibinfo {year}
  {2009})%
  \bibAnnoteFile{NoStop}{Demir:2008tr}%
\bibitem{Bozek:2010er}%
  \BibitemOpen
  \bibfield{author}{%
  \bibinfo {author} {\bibfnamefont{P.}~\bibnamefont{Bo\.zek}},\ }%
  \bibfield{journal}{%
  \Doi{10.1103/PhysRevC.83.044910}{\bibinfo {journal} {Phys. Rev.}}\ }%
  \textbf{\bibinfo {volume} {C83}},\ \bibinfo {pages} {044910} (\bibinfo {year}
  {2011})%
  \bibAnnoteFile{NoStop}{Bozek:2010er}%
\bibitem{Bozek:2011wa}%
  \BibitemOpen
  \bibfield{author}{%
  \bibinfo {author} {\bibfnamefont{P.}~\bibnamefont{Bo\.zek}},\ }%
  \bibfield{journal}{%
  \Doi{10.1016/j.physletb.2011.04.020}{\bibinfo {journal} {Phys. Lett.}}\ }%
  \textbf{\bibinfo {volume} {B699}},\ \bibinfo {pages} {283} (\bibinfo {year}
  {2011})%
  \bibAnnoteFile{NoStop}{Bozek:2011wa}%
\bibitem{Chojnacki:2007jc}%
  \BibitemOpen
  \bibfield{author}{%
  \bibinfo {author} {\bibfnamefont{M.}~\bibnamefont{Chojnacki}}\ and\ \bibinfo
  {author} {\bibfnamefont{W.}~\bibnamefont{Florkowski}},\ }%
  \bibfield{journal}{%
  \bibinfo {journal} {Acta Phys. Polon.}\ }%
  \textbf{\bibinfo {volume} {B38}},\ \bibinfo {pages} {3249} (\bibinfo {year}
  {2007})%
  \bibAnnoteFile{NoStop}{Chojnacki:2007jc}%
\bibitem{Broniowski:2008vp}%
  \BibitemOpen
  \bibfield{author}{%
  \bibinfo {author} {\bibfnamefont{W.}~\bibnamefont{Broniowski}}, \bibinfo
  {author} {\bibfnamefont{M.}~\bibnamefont{Chojnacki}}, \bibinfo {author}
  {\bibfnamefont{W.}~\bibnamefont{Florkowski}},\ and\ \bibinfo {author}
  {\bibfnamefont{A.}~\bibnamefont{Kisiel}},\ }%
  \bibfield{journal}{%
  \Doi{10.1103/PhysRevLett.101.022301}{\bibinfo {journal} {Phys. Rev. Lett.}}\
  }%
  \textbf{\bibinfo {volume} {101}},\ \bibinfo {pages} {022301} (\bibinfo {year}
  {2008})%
  \bibAnnoteFile{NoStop}{Broniowski:2008vp}%
\bibitem{Pratt:2008qv}%
  \BibitemOpen
  \bibfield{author}{%
  \bibinfo {author} {\bibfnamefont{S.}~\bibnamefont{Pratt}},\ }%
  \bibfield{journal}{%
  \Doi{10.1103/PhysRevLett.102.232301}{\bibinfo {journal} {Phys. Rev. Lett.}}\
  }%
  \textbf{\bibinfo {volume} {102}},\ \bibinfo {pages} {232301} (\bibinfo {year}
  {2009})%
  \bibAnnoteFile{NoStop}{Pratt:2008qv}%
\bibitem{Huovinen:2009yb}%
  \BibitemOpen
  \bibfield{author}{%
  \bibinfo {author} {\bibfnamefont{P.}~\bibnamefont{Huovinen}}\ and\ \bibinfo
  {author} {\bibfnamefont{P.}~\bibnamefont{Petreczky}},\ }%
  \bibfield{journal}{%
  \Doi{10.1016/j.nuclphysa.2010.02.015}{\bibinfo {journal} {Nucl.Phys.}}\ }%
  \textbf{\bibinfo {volume} {A837}},\ \bibinfo {pages} {26} (\bibinfo {year}
  {2010})%
  \bibAnnoteFile{NoStop}{Huovinen:2009yb}%
\bibitem{Borsanyi:2010cj}%
  \BibitemOpen
  \bibfield{author}{%
  \bibinfo {author} {\bibfnamefont{S.}~\bibnamefont{Borsanyi}} \emph{et~al.},\
  }%
  \bibfield{journal}{%
  \Doi{10.1007/JHEP11(2010)077}{\bibinfo {journal} {JHEP}}\ }%
  \textbf{\bibinfo {volume} {11}},\ \bibinfo {pages} {077} (\bibinfo {year}
  {2010})%
  \bibAnnoteFile{NoStop}{Borsanyi:2010cj}%
\bibitem{Bialas:2004su}%
  \BibitemOpen
  \bibfield{author}{%
  \bibinfo {author} {\bibfnamefont{A.}~\bibnamefont{Bia\l{}as}}\ and\ \bibinfo
  {author} {\bibfnamefont{W.}~\bibnamefont{Czy\.z}},\ }%
  \bibfield{journal}{%
  \bibinfo {journal} {Acta Phys. Polon.}\ }%
  \textbf{\bibinfo {volume} {B36}},\ \bibinfo {pages} {905} (\bibinfo {year}
  {2005})%
  \bibAnnoteFile{NoStop}{Bialas:2004su}%
\bibitem{Bozek:2010bi}%
  \BibitemOpen
  \bibfield{author}{%
  \bibinfo {author} {\bibfnamefont{P.}~\bibnamefont{Bo\.zek}}\ and\ \bibinfo
  {author} {\bibfnamefont{I.}~\bibnamefont{Wyskiel}},\ }%
  \bibfield{journal}{%
  \Doi{10.1103/PhysRevC.81.054902}{\bibinfo {journal} {Phys. Rev.}}\ }%
  \textbf{\bibinfo {volume} {C81}},\ \bibinfo {pages} {054902} (\bibinfo {year}
  {2010})%
  \bibAnnoteFile{NoStop}{Bozek:2010bi}%
\bibitem{Alver:2008zz}%
  \BibitemOpen
  \bibfield{author}{%
  \bibinfo {author} {\bibfnamefont{B.}~\bibnamefont{Alver}} \emph{et~al.},\ }%
  \bibfield{journal}{%
  \Doi{10.1103/PhysRevC.77.014906}{\bibinfo {journal} {Phys. Rev.}}\ }%
  \textbf{\bibinfo {volume} {C77}},\ \bibinfo {pages} {014906} (\bibinfo {year}
  {2008})%
  \bibAnnoteFile{NoStop}{Alver:2008zz}%
\bibitem{Broniowski:2007nz}%
  \BibitemOpen
  \bibfield{author}{%
  \bibinfo {author} {\bibfnamefont{W.}~\bibnamefont{Broniowski}}, \bibinfo
  {author} {\bibfnamefont{M.}~\bibnamefont{Rybczy\'nski}},\ and\ \bibinfo
  {author} {\bibfnamefont{P.}~\bibnamefont{Bo\.zek}},\ }%
  \bibfield{journal}{%
  \Doi{10.1016/j.cpc.2008.07.016}{\bibinfo {journal} {Comput. Phys. Commun.}}\
  }%
  \textbf{\bibinfo {volume} {180}},\ \bibinfo {pages} {69} (\bibinfo {year}
  {2009})%
  \bibAnnoteFile{NoStop}{Broniowski:2007nz}%
\bibitem{Huovinen:2008te}%
  \BibitemOpen
  \bibfield{author}{%
  \bibinfo {author} {\bibfnamefont{P.}~\bibnamefont{Huovinen}}\ and\ \bibinfo
  {author} {\bibfnamefont{D.}~\bibnamefont{Molnar}},\ }%
  \bibfield{journal}{%
  \Doi{10.1103/PhysRevC.79.014906}{\bibinfo {journal} {Phys. Rev.}}\ }%
  \textbf{\bibinfo {volume} {C79}},\ \bibinfo {pages} {014906} (\bibinfo {year}
  {2009})%
  \bibAnnoteFile{NoStop}{Huovinen:2008te}%
\bibitem{Gavin:1985ph}%
  \BibitemOpen
  \bibfield{author}{%
  \bibinfo {author} {\bibfnamefont{S.}~\bibnamefont{Gavin}},\ }%
  \bibfield{journal}{%
  \Doi{10.1016/0375-9474(85)90190-3}{\bibinfo {journal} {Nucl. Phys.}}\ }%
  \textbf{\bibinfo {volume} {A435}},\ \bibinfo {pages} {826} (\bibinfo {year}
  {1985})%
  \bibAnnoteFile{NoStop}{Gavin:1985ph}%
\bibitem{Hosoya:1983xm}%
  \BibitemOpen
  \bibfield{author}{%
  \bibinfo {author} {\bibfnamefont{A.}~\bibnamefont{Hosoya}}\ and\ \bibinfo
  {author} {\bibfnamefont{K.}~\bibnamefont{Kajantie}},\ }%
  \bibfield{journal}{%
  \Doi{10.1016/0550-3213(85)90499-7}{\bibinfo {journal} {Nucl. Phys.}}\ }%
  \textbf{\bibinfo {volume} {B250}},\ \bibinfo {pages} {666} (\bibinfo {year}
  {1985})%
  \bibAnnoteFile{NoStop}{Hosoya:1983xm}%
\bibitem{Sasaki:2008fg}%
  \BibitemOpen
  \bibfield{author}{%
  \bibinfo {author} {\bibfnamefont{C.}~\bibnamefont{Sasaki}}\ and\ \bibinfo
  {author} {\bibfnamefont{K.}~\bibnamefont{Redlich}},\ }%
  \bibfield{journal}{%
  \Doi{10.1103/PhysRevC.79.055207}{\bibinfo {journal} {Phys. Rev.}}\ }%
  \textbf{\bibinfo {volume} {C79}},\ \bibinfo {pages} {055207} (\bibinfo {year}
  {2009})%
  \bibAnnoteFile{NoStop}{Sasaki:2008fg}%
\bibitem{Teaney:2003kp}%
  \BibitemOpen
  \bibfield{author}{%
  \bibinfo {author} {\bibfnamefont{D.}~\bibnamefont{Teaney}},\ }%
  \bibfield{journal}{%
  \bibinfo {journal} {Phys. Rev.}\ }%
  \textbf{\bibinfo {volume} {C68}},\ \bibinfo {pages} {034913} (\bibinfo {year}
  {2003})%
  \bibAnnoteFile{NoStop}{Teaney:2003kp}%
\bibitem{Monnai:2009ad}%
  \BibitemOpen
  \bibfield{author}{%
  \bibinfo {author} {\bibfnamefont{A.}~\bibnamefont{Monnai}}\ and\ \bibinfo
  {author} {\bibfnamefont{T.}~\bibnamefont{Hirano}},\ }%
  \bibfield{journal}{%
  \Doi{10.1103/PhysRevC.80.054906}{\bibinfo {journal} {Phys. Rev.}}\ }%
  \textbf{\bibinfo {volume} {C80}},\ \bibinfo {pages} {054906} (\bibinfo {year}
  {2009})%
  \bibAnnoteFile{NoStop}{Monnai:2009ad}%
\bibitem{Pratt:2010jt}%
  \BibitemOpen
  \bibfield{author}{%
  \bibinfo {author} {\bibfnamefont{S.}~\bibnamefont{Pratt}}\ and\ \bibinfo
  {author} {\bibfnamefont{G.}~\bibnamefont{Torrieri}},\ }%
  \bibfield{journal}{%
  \Doi{10.1103/PhysRevC.82.044901}{\bibinfo {journal} {Phys.Rev.}}\ }%
  \textbf{\bibinfo {volume} {C82}},\ \bibinfo {pages} {044901} (\bibinfo {year}
  {2010})%
  \bibAnnoteFile{NoStop}{Pratt:2010jt}%
\bibitem{Dusling:2011fd}%
  \BibitemOpen
  \bibfield{author}{%
  \bibinfo {author} {\bibfnamefont{K.}~\bibnamefont{Dusling}}\ and\ \bibinfo
  {author} {\bibfnamefont{T.}~\bibnamefont{Schafer}}}%
   (\bibinfo {year} {2011}),\
  \Eprint{http://arxiv.org/abs/1109.5181}{arXiv:1109.5181 [hep-ph]}%
  \bibAnnoteFile{NoStop}{Dusling:2011fd}%
\bibitem{Dusling:2009df}%
  \BibitemOpen
  \bibfield{author}{%
  \bibinfo {author} {\bibfnamefont{K.}~\bibnamefont{Dusling}}, \bibinfo
  {author} {\bibfnamefont{G.~D.}\ \bibnamefont{Moore}},\ and\ \bibinfo {author}
  {\bibfnamefont{D.}~\bibnamefont{Teaney}},\ }%
  \bibfield{journal}{%
  \Doi{10.1103/PhysRevC.81.034907}{\bibinfo {journal} {Phys. Rev.}}\ }%
  \textbf{\bibinfo {volume} {C81}},\ \bibinfo {pages} {034907} (\bibinfo {year}
  {2010})%
  \bibAnnoteFile{NoStop}{Dusling:2009df}%
\bibitem{Molnar:2011kx}%
  \BibitemOpen
  \bibfield{author}{%
  \bibinfo {author} {\bibfnamefont{D.}~\bibnamefont{Molnar}},\ }%
  \bibfield{journal}{%
  \Doi{10.1088/0954-3899/38/12/124173}{\bibinfo {journal} {J.Phys.G}}\ }%
  \textbf{\bibinfo {volume} {G38}},\ \bibinfo {pages} {124173} (\bibinfo {year}
  {2011})%
  \bibAnnoteFile{NoStop}{Molnar:2011kx}%
\bibitem{Bozek:2011ph}%
  \BibitemOpen
  \bibfield{author}{%
  \bibinfo {author} {\bibfnamefont{P.}~\bibnamefont{Bozek}},\ }%
  \bibfield{journal}{%
  \Doi{10.1088/0954-3899/38/12/124043}{\bibinfo {journal} {J.Phys.G}}\ }%
  \textbf{\bibinfo {volume} {G38}},\ \bibinfo {pages} {124043} (\bibinfo {year}
  {2011})%
  \bibAnnoteFile{NoStop}{Bozek:2011ph}%
\bibitem{Chojnacki:2011hb}%
  \BibitemOpen
  \bibfield{author}{%
  \bibinfo {author} {\bibfnamefont{M.}~\bibnamefont{Chojnacki}}, \bibinfo
  {author} {\bibfnamefont{A.}~\bibnamefont{Kisiel}}, \bibinfo {author}
  {\bibfnamefont{W.}~\bibnamefont{Florkowski}},\ and\ \bibinfo {author}
  {\bibfnamefont{W.}~\bibnamefont{Broniowski}},\ }%
  \bibfield{journal}{%
  \Doi{10.1016/j.cpc.2011.11.018}{\bibinfo {journal} {Comput.Phys.Commun.}}\ }%
  \textbf{\bibinfo {volume} {183}},\ \bibinfo {pages} {746} (\bibinfo {year}
  {2012})%
  \bibAnnoteFile{NoStop}{Chojnacki:2011hb}%
\bibitem{Andronic:2005yp}%
  \BibitemOpen
  \bibfield{author}{%
  \bibinfo {author} {\bibfnamefont{A.}~\bibnamefont{Andronic}}, \bibinfo
  {author} {\bibfnamefont{P.}~\bibnamefont{Braun-Munzinger}},\ and\ \bibinfo
  {author} {\bibfnamefont{J.}~\bibnamefont{Stachel}},\ }%
  \bibfield{journal}{%
  \Doi{10.1016/j.nuclphysa.2006.03.012}{\bibinfo {journal} {Nucl. Phys.}}\ }%
  \textbf{\bibinfo {volume} {A772}},\ \bibinfo {pages} {167} (\bibinfo {year}
  {2006})%
  \bibAnnoteFile{NoStop}{Andronic:2005yp}%
\bibitem{Bluhm:2007nu}%
  \BibitemOpen
  \bibfield{author}{%
  \bibinfo {author} {\bibfnamefont{M.}~\bibnamefont{Bluhm}}, \bibinfo {author}
  {\bibfnamefont{B.}~\bibnamefont{Kampfer}}, \bibinfo {author}
  {\bibfnamefont{R.}~\bibnamefont{Schulze}}, \bibinfo {author}
  {\bibfnamefont{D.}~\bibnamefont{Seipt}},\ and\ \bibinfo {author}
  {\bibfnamefont{U.}~\bibnamefont{Heinz}},\ }%
  \bibfield{journal}{%
  \Doi{10.1103/PhysRevC.76.034901}{\bibinfo {journal} {Phys. Rev.}}\ }%
  \textbf{\bibinfo {volume} {C76}},\ \bibinfo {pages} {034901} (\bibinfo {year}
  {2007})%
  \bibAnnoteFile{NoStop}{Bluhm:2007nu}%
\bibitem{Schenke:2011bn}%
  \BibitemOpen
  \bibfield{author}{%
  \bibinfo {author} {\bibfnamefont{B.}~\bibnamefont{Schenke}}, \bibinfo
  {author} {\bibfnamefont{S.}~\bibnamefont{Jeon}},\ and\ \bibinfo {author}
  {\bibfnamefont{C.}~\bibnamefont{Gale}}}%
   (\bibinfo {year} {2011}),\
  \Eprint{http://arxiv.org/abs/1109.6289}{arXiv:1109.6289 [hep-ph]}%
  \bibAnnoteFile{NoStop}{Schenke:2011bn}%
\bibitem{Bozek:2007qt}%
  \BibitemOpen
  \bibfield{author}{%
  \bibinfo {author} {\bibfnamefont{P.}~\bibnamefont{Bo\.zek}},\ }%
  \bibfield{journal}{%
  \Doi{10.1103/PhysRevC.77.034911}{\bibinfo {journal} {Phys. Rev.}}\ }%
  \textbf{\bibinfo {volume} {C77}},\ \bibinfo {pages} {034911} (\bibinfo {year}
  {2008})%
  \bibAnnoteFile{NoStop}{Bozek:2007qt}%
\bibitem{Monnai:2011ju}%
  \BibitemOpen
  \bibfield{author}{%
  \bibinfo {author} {\bibfnamefont{A.}~\bibnamefont{Monnai}}\ and\ \bibinfo
  {author} {\bibfnamefont{T.}~\bibnamefont{Hirano}},\ }%
  \bibfield{journal}{%
  \Doi{10.1016/j.physletb.2011.08.049}{\bibinfo {journal} {Phys.Lett.}}\ }%
  \textbf{\bibinfo {volume} {B703}},\ \bibinfo {pages} {583} (\bibinfo {year}
  {2011})%
  \bibAnnoteFile{NoStop}{Monnai:2011ju}%
\bibitem{Back:2002wb}%
  \BibitemOpen
  \bibfield{author}{%
  \bibinfo {author} {\bibfnamefont{B.~B.}\ \bibnamefont{Back}} \emph{et~al.},\
  }%
  \bibfield{journal}{%
  \Doi{10.1103/PhysRevLett.91.052303}{\bibinfo {journal} {Phys. Rev. Lett.}}\
  }%
  \textbf{\bibinfo {volume} {91}},\ \bibinfo {pages} {052303} (\bibinfo {year}
  {2003})%
  \bibAnnoteFile{NoStop}{Back:2002wb}%
\bibitem{Bearden:2001qq}%
  \BibitemOpen
  \bibfield{author}{%
  \bibinfo {author} {\bibfnamefont{I.~G.}\ \bibnamefont{Bearden}} \emph{et~al.}
  (\bibinfo {collaboration} {BRAHMS}),\ }%
  \bibfield{journal}{%
  \Doi{10.1103/PhysRevLett.88.202301}{\bibinfo {journal} {Phys. Rev. Lett.}}\
  }%
  \textbf{\bibinfo {volume} {88}},\ \bibinfo {pages} {202301} (\bibinfo {year}
  {2002})%
  \bibAnnoteFile{NoStop}{Bearden:2001qq}%
\bibitem{Adler:2003cb}%
  \BibitemOpen
  \bibfield{author}{%
  \bibinfo {author} {\bibfnamefont{S.~S.}\ \bibnamefont{Adler}} \emph{et~al.}
  (\bibinfo {collaboration} {PHENIX}),\ }%
  \bibfield{journal}{%
  \Doi{10.1103/PhysRevC.69.034909}{\bibinfo {journal} {Phys. Rev.}}\ }%
  \textbf{\bibinfo {volume} {C69}},\ \bibinfo {pages} {034909} (\bibinfo {year}
  {2004})%
  \bibAnnoteFile{NoStop}{Adler:2003cb}%
\bibitem{Back:2004dy}%
  \BibitemOpen
  \bibfield{author}{%
  \bibinfo {author} {\bibfnamefont{B.~B.}\ \bibnamefont{Back}} \emph{et~al.}
  (\bibinfo {collaboration} {PHOBOS}),\ }%
  \bibfield{journal}{%
  \bibinfo {journal} {Phys. Rev.}\ }%
  \textbf{\bibinfo {volume} {C70}},\ \bibinfo {pages} {021902} (\bibinfo {year}
  {2004})%
  \bibAnnoteFile{NoStop}{Back:2004dy}%
\bibitem{Back:2004mh}%
  \BibitemOpen
  \bibfield{author}{%
  \bibinfo {author} {\bibfnamefont{B.~B.}\ \bibnamefont{Back}} \emph{et~al.}
  (\bibinfo {collaboration} {PHOBOS}),\ }%
  \bibfield{journal}{%
  \Doi{10.1103/PhysRevC.72.051901}{\bibinfo {journal} {Phys. Rev.}}\ }%
  \textbf{\bibinfo {volume} {C72}},\ \bibinfo {pages} {051901} (\bibinfo {year}
  {2005})%
  \bibAnnoteFile{NoStop}{Back:2004mh}%
\bibitem{Cleymans:2004pp}%
  \BibitemOpen
  \bibfield{author}{%
  \bibinfo {author} {\bibfnamefont{J.}~\bibnamefont{Cleymans}}, \bibinfo
  {author} {\bibfnamefont{B.}~\bibnamefont{Kampfer}}, \bibinfo {author}
  {\bibfnamefont{M.}~\bibnamefont{Kaneta}}, \bibinfo {author}
  {\bibfnamefont{S.}~\bibnamefont{Wheaton}},\ and\ \bibinfo {author}
  {\bibfnamefont{N.}~\bibnamefont{Xu}},\ }%
  \bibfield{journal}{%
  \Doi{10.1103/PhysRevC.71.054901}{\bibinfo {journal} {Phys. Rev.}}\ }%
  \textbf{\bibinfo {volume} {C71}},\ \bibinfo {pages} {054901} (\bibinfo {year}
  {2005})%
  \bibAnnoteFile{NoStop}{Cleymans:2004pp}%
\bibitem{Bozek:2005eu}%
  \BibitemOpen
  \bibfield{author}{%
  \bibinfo {author} {\bibfnamefont{P.}~\bibnamefont{Bo\.zek}},\ }%
  \bibfield{journal}{%
  \bibinfo {journal} {Acta Phys. Polon.}\ }%
  \textbf{\bibinfo {volume} {B36}},\ \bibinfo {pages} {3071} (\bibinfo {year}
  {2005})%
  \bibAnnoteFile{NoStop}{Bozek:2005eu}%
\bibitem{Becattini:2008ya}%
  \BibitemOpen
  \bibfield{author}{%
  \bibinfo {author} {\bibfnamefont{F.}~\bibnamefont{Becattini}}\ and\ \bibinfo
  {author} {\bibfnamefont{J.}~\bibnamefont{Manninen}},\ }%
  \bibfield{journal}{%
  \Doi{10.1016/j.physletb.2009.01.066}{\bibinfo {journal} {Phys. Lett.}}\ }%
  \textbf{\bibinfo {volume} {B673}},\ \bibinfo {pages} {19} (\bibinfo {year}
  {2009})%
  \bibAnnoteFile{NoStop}{Becattini:2008ya}%
\bibitem{Adare:2011tg}%
  \BibitemOpen
  \bibfield{author}{%
  \bibinfo {author} {\bibfnamefont{A.}~\bibnamefont{Adare}} \emph{et~al.}
  (\bibinfo {collaboration} {PHENIX Collaboration}),\ }%
  \bibfield{journal}{%
  \Doi{10.1103/PhysRevLett.107.252301}{\bibinfo {journal} {Phys.Rev.Lett.}}\ }%
  \textbf{\bibinfo {volume} {107}},\ \bibinfo {pages} {252301} (\bibinfo {year}
  {2011})%
  \bibAnnoteFile{NoStop}{Adare:2011tg}%
\bibitem{Back:2005pc}%
  \BibitemOpen
  \bibfield{author}{%
  \bibinfo {author} {\bibfnamefont{B.~B.}\ \bibnamefont{Back}} \emph{et~al.}
  (\bibinfo {collaboration} {PHOBOS}),\ }%
  \bibfield{journal}{%
  \Doi{10.1103/PhysRevLett.97.012301}{\bibinfo {journal} {Phys. Rev. Lett.}}\
  }%
  \textbf{\bibinfo {volume} {97}},\ \bibinfo {pages} {012301} (\bibinfo {year}
  {2006})%
  \bibAnnoteFile{NoStop}{Back:2005pc}%
\bibitem{Abelev:2008jga}%
  \BibitemOpen
  \bibfield{author}{%
  \bibinfo {author} {\bibfnamefont{B.~I.}\ \bibnamefont{Abelev}} \emph{et~al.}
  (\bibinfo {collaboration} {STAR}),\ }%
  \bibfield{journal}{%
  \Doi{10.1103/PhysRevLett.101.252301}{\bibinfo {journal} {Phys. Rev. Lett.}}\
  }%
  \textbf{\bibinfo {volume} {101}},\ \bibinfo {pages} {252301} (\bibinfo {year}
  {2008})%
  \bibAnnoteFile{NoStop}{Abelev:2008jga}%
\bibitem{Bozek:2009mz}%
  \BibitemOpen
  \bibfield{author}{%
  \bibinfo {author} {\bibfnamefont{P.}~\bibnamefont{Bo\.zek}}\ and\ \bibinfo
  {author} {\bibfnamefont{I.}~\bibnamefont{Wyskiel}},\ }%
  \bibfield{journal}{%
  \bibinfo {journal} {PoS}\ }%
  \textbf{\bibinfo {volume} {EPS-HEP 2009}},\ \bibinfo {pages} {039} (\bibinfo
  {year} {2009})%
  \bibAnnoteFile{NoStop}{Bozek:2009mz}%
\bibitem{Qiu:2011fi}%
  \BibitemOpen
  \bibfield{author}{%
  \bibinfo {author} {\bibfnamefont{Z.}~\bibnamefont{Qiu}}\ and\ \bibinfo
  {author} {\bibfnamefont{U.~W.}\ \bibnamefont{Heinz}}}%
   (\bibinfo {year} {2011}),\
  \Eprint{http://arxiv.org/abs/1108.1714}{arXiv:1108.1714 [nucl-th]}%
  \bibAnnoteFile{NoStop}{Qiu:2011fi}%
\bibitem{Adams:2004yc}%
  \BibitemOpen
  \bibfield{author}{%
  \bibinfo {author} {\bibfnamefont{J.}~\bibnamefont{Adams}} \emph{et~al.}
  (\bibinfo {collaboration} {STAR}),\ }%
  \bibfield{journal}{%
  \bibinfo {journal} {Phys. Rev.}\ }%
  \textbf{\bibinfo {volume} {C71}},\ \bibinfo {pages} {044906} (\bibinfo {year}
  {2005})%
  \bibAnnoteFile{NoStop}{Adams:2004yc}%
\bibitem{Bozek:2010aj}%
  \BibitemOpen
  \bibfield{author}{%
  \bibinfo {author} {\bibfnamefont{P.}~\bibnamefont{Bo\.zek}}\ and\ \bibinfo
  {author} {\bibfnamefont{I.}~\bibnamefont{Wyskiel-Piekarska}},\ }%
  \bibfield{journal}{%
  \Doi{10.1103/PhysRevC.83.024910}{\bibinfo {journal} {Phys.Rev.}}\ }%
  \textbf{\bibinfo {volume} {C83}},\ \bibinfo {pages} {024910} (\bibinfo {year}
  {2011})%
  \bibAnnoteFile{NoStop}{Bozek:2010aj}%
\bibitem{Bozek:2007di}%
  \BibitemOpen
  \bibfield{author}{%
  \bibinfo {author} {\bibfnamefont{P.}~\bibnamefont{Bo\.zek}},\ }%
  \bibfield{journal}{%
  \bibinfo {journal} {Acta Phys. Polon.}\ }%
  \textbf{\bibinfo {volume} {B39}},\ \bibinfo {pages} {1375} (\bibinfo {year}
  {2008})%
  \bibAnnoteFile{NoStop}{Bozek:2007di}%
\bibitem{Ryblewski:2010bs}%
  \BibitemOpen
  \bibfield{author}{%
  \bibinfo {author} {\bibfnamefont{R.}~\bibnamefont{Ryblewski}}\ and\ \bibinfo
  {author} {\bibfnamefont{W.}~\bibnamefont{Florkowski}},\ }%
  \bibfield{journal}{%
  \Doi{10.1088/0954-3899/38/1/015104}{\bibinfo {journal} {J. Phys.}}\ }%
  \textbf{\bibinfo {volume} {G38}},\ \bibinfo {pages} {015104} (\bibinfo {year}
  {2011})%
  \bibAnnoteFile{NoStop}{Ryblewski:2010bs}%
\bibitem{Martinez:2010sc}%
  \BibitemOpen
  \bibfield{author}{%
  \bibinfo {author} {\bibfnamefont{M.}~\bibnamefont{Martinez}}\ and\ \bibinfo
  {author} {\bibfnamefont{M.}~\bibnamefont{Strickland}},\ }%
  \bibfield{journal}{%
  \Doi{10.1016/j.nuclphysa.2010.08.011}{\bibinfo {journal} {Nucl. Phys.}}\ }%
  \textbf{\bibinfo {volume} {A848}},\ \bibinfo {pages} {183} (\bibinfo {year}
  {2010})%
  \bibAnnoteFile{NoStop}{Martinez:2010sc}%
\bibitem{Beuf:2009cx}%
  \BibitemOpen
  \bibfield{author}{%
  \bibinfo {author} {\bibfnamefont{G.}~\bibnamefont{Beuf}}, \bibinfo {author}
  {\bibfnamefont{M.~P.}\ \bibnamefont{Heller}}, \bibinfo {author}
  {\bibfnamefont{R.~A.}\ \bibnamefont{Janik}},\ and\ \bibinfo {author}
  {\bibfnamefont{R.}~\bibnamefont{Peschanski}},\ }%
  \bibfield{journal}{%
  \Doi{10.1088/1126-6708/2009/10/043}{\bibinfo {journal} {JHEP}}\ }%
  \textbf{\bibinfo {volume} {10}},\ \bibinfo {pages} {043} (\bibinfo {year}
  {2009})%
  \bibAnnoteFile{NoStop}{Beuf:2009cx}%
\bibitem{Heller:2011ju}%
  \BibitemOpen
  \bibfield{author}{%
  \bibinfo {author} {\bibfnamefont{M.~P.}\ \bibnamefont{Heller}}, \bibinfo
  {author} {\bibfnamefont{R.~A.}\ \bibnamefont{Janik}},\ and\ \bibinfo {author}
  {\bibfnamefont{P.}~\bibnamefont{Witaszczyk}}}%
   (\bibinfo {year} {2011}),\
  \Eprint{http://arxiv.org/abs/1103.3452}{arXiv:1103.3452 [hep-th]}%
  \bibAnnoteFile{NoStop}{Heller:2011ju}%
\bibitem{Bozek:2009ty}%
  \BibitemOpen
  \bibfield{author}{%
  \bibinfo {author} {\bibfnamefont{P.}~\bibnamefont{Bo\.zek}}\ and\ \bibinfo
  {author} {\bibfnamefont{I.}~\bibnamefont{Wyskiel}},\ }%
  \bibfield{journal}{%
  \Doi{10.1103/PhysRevC.79.044916}{\bibinfo {journal} {Phys. Rev.}}\ }%
  \textbf{\bibinfo {volume} {C79}},\ \bibinfo {pages} {044916} (\bibinfo {year}
  {2009})%
  \bibAnnoteFile{NoStop}{Bozek:2009ty}%
\end{thebibliography}%


\end{document}